\documentclass[letterpaper,english,aps,prb,floatfix,showpacs,twocolumn,amsfonts,amssymb,superscriptaddress]{revtex4-1}

\usepackage{balance}
\usepackage[T1]{fontenc}
\usepackage[applemac]{inputenc}
\usepackage[english]{babel}
\usepackage{graphics}
\usepackage{amssymb}
\usepackage{amsmath}
\usepackage{overpic}
\usepackage[toc]{appendix}

\usepackage{natbib}
\usepackage{xcolor}
\usepackage{soul}
\usepackage{commath}
\usepackage{tikz}
\usepackage{physics}

\usepackage{hyperref}

\newcommand{\be}{\begin{equation}}
\newcommand{\ee}{\end{equation}}
\newcommand{\bspl}{\begin{split}}
\newcommand{\espl}{\end{split}}
\newcommand{\bea}{\begin{eqnarray}}
\newcommand{\eea}{\end{eqnarray}}

\newcommand{\bd}{\boldsymbol}
\newcommand{\eq}{\equiv}
\newcommand{\cev}[1]{\reflectbox{\ensuremath{\vec{\reflectbox{\ensuremath{#1}}}}}}

\def\a{\alpha}
\def\b{\beta}
\def\e{\varepsilon}
\def\d{\delta}
\def\g{\gamma}
\def\k{\kappa}
\def\m{\mu}
\def\l{\lambda}
\def\th{\theta}
\def\t{\tau}

\def\o{\omega}

\def\s{\sigma}

\def\ct{\tilde c}

\def\D{\Delta}
\def\O{\Omega}
\def\L{\Lambda}
\def\S{\Sigma}

\def\ra{\rightarrow}
\def\up{\uparrow}
\def\pll{\parallel}
\def\down{\downarrow}

\def\pd{\partial}
\def\nb{\nabla}
\def\bdnb{\bd{\nb}}
\def\bk{{\bf k}}

\def\bq{{\bf q}}
\def\br{{\bf r}}

\def\bA{{\bf A}}
\def\bB{{\bf B}}
\def\bE{{\bf E}}

\def\bJ{{\bf J}}

\def\bx{{\bf x}}
\def\by{{\bf y}}
\def\bz{{\bf z}}
\def\bo{{\bf 0}}
\def\bpsi{\bd{\psi}}

\def\nn{\nonumber}
\def\lb{\label}
\def\pref#1{(\ref{#1})}

\newcount\bozza \bozza=0
\ifnum\bozza=1
\newdimen\shift \shift=-2truecm
\def\lb#1{%
{\label{#1}\rlap{\kern\shift{$\scriptstyle#1$}}}}
\else\def\lb#1{\label{#1}} \fi

\definecolor{darkred}{rgb}{0.55, 0.0, 0.0}
\definecolor{darkpowderblue}{rgb}{0.0, 0.2, 0.6}

\usetikzlibrary{shapes.arrows, fadings}
\tikzfading[name=fadeleft,
  left color=transparent!80, right color=transparent!0]

\begin{document}
\title{Generalized plasma waves in layered superconductors}
\author{F. Gabriele}
\affiliation{Department of Physics and ISC-CNR, ``Sapienza'' University of Rome, P.le A. Moro 5, 00185 Rome, Italy}
\author{C. Castellani}
\affiliation{Department of Physics and ISC-CNR, ``Sapienza'' University of Rome, P.le A. Moro 5, 00185 Rome, Italy}
\author{L. Benfatto}
 \email{lara.benfatto@roma1.infn.it}
\affiliation{Department of Physics and ISC-CNR, ``Sapienza'' University of Rome, P.le A. Moro 5, 00185 Rome, Italy}

\begin{abstract}
In a layered and strongly anisotropic superconductor the hybrid modes provided by the propagation of electromagnetic waves in the matter identify two well separate energy scales connected to the large in-plane plasma frequency  and to the soft out-of-plane Josephson plasmon. Despite the wide interest in their detection and manipulation by means of different experimental protocols, a unified description of plasma waves valid at arbitrary energy and momentum is still lacking. Here we provide a complete description of generalized plasma waves in a layered superconductors by taking advantage of their connection to the gauge-invariant superconducting phase. 
We show that the anisotropy of the superfluid response leads to two intertwined hybrid light-matter modes with mixed longitudinal and transverse character, while a purely longitudinal plasmon is only recovered for wavevectors larger than the crossover scale set in by the plasma-frequencies anisotropy. Interestingly, below such scale both modes appears with equal weight in the physical density response. Our results open a promising perspective for plasmonic applications made possible by the next-generation spectroscopic techniques able to combine sub-micron momentum resolution with THz energy resolution. 

\end{abstract}
\date{\today}

\maketitle

\section{Introduction}

Plasmons represent the fundamental excitations of the conduction electrons in metals, and their existence can be easily understood within a classical framework based on Maxwell's equations. Indeed, in a source-free metal bulk plasmons are characterized by zero magnetic field and longitudinal electric field ($\nb \times \bE=0$), so that they trivially satisfy Maxwell's equations under the condition of vanishing permittivity\cite{maier}.  Since the longitudinal electric field couples to density fluctuations the plasma excitation appears also in the charge-density response as collective modes of the electron gas\cite{pines,fetter}.  Such a hybrid light-matter mode is also named plasma polariton\cite{koppens-review-polaritons}, especially in the context where spatial confinement at the interfaces between a thin metallic film and a dielectric gives rise to a propagating mode, that can be observed with scanning near-field optical microscopy\cite{basov-review-polaritons}. When the metal undergoes a superconducting (SC) transitions plasmons characterize also the fluctuation spectrum of the superconducting phase of the complex order parameter formed below the SC temperature $T_c$\cite{nagaosa}. Indeed, as originally pointed out by Anderson\cite{anderson_pr58}, the sound-like propagating phase mode of the neutral superfluid\cite{pines} is converted into gapped plasma oscillations in a charged superconductor. The appearance of the plasma mode in the spectrum of phase fluctuations is a natural consequence of the fact that the quantum phase of electrons is the variable conjugate to the density\cite{anderson_pr58,aitchison_prb95,depalo_prb99,nagaosa}. From the theoretically point of view, such an equivalence has been often exploited in the literature to determine the plasma dispersion of a superconductor via the study of the spectrum of the phase mode. The latter can be carried out in a rather elegant and compact way by deriving directly the quantum action for the phase degrees of freedom\cite{nagaosa,aitchison_prb95,depalo_prb99,randeria_prb00,benfatto_prb01,benfatto_prb04,millis_prr20}, as we will discuss in details in this manuscript.  

\begin{figure}[ht]
    \centering
       \includegraphics[width=0.5\textwidth,keepaspectratio]{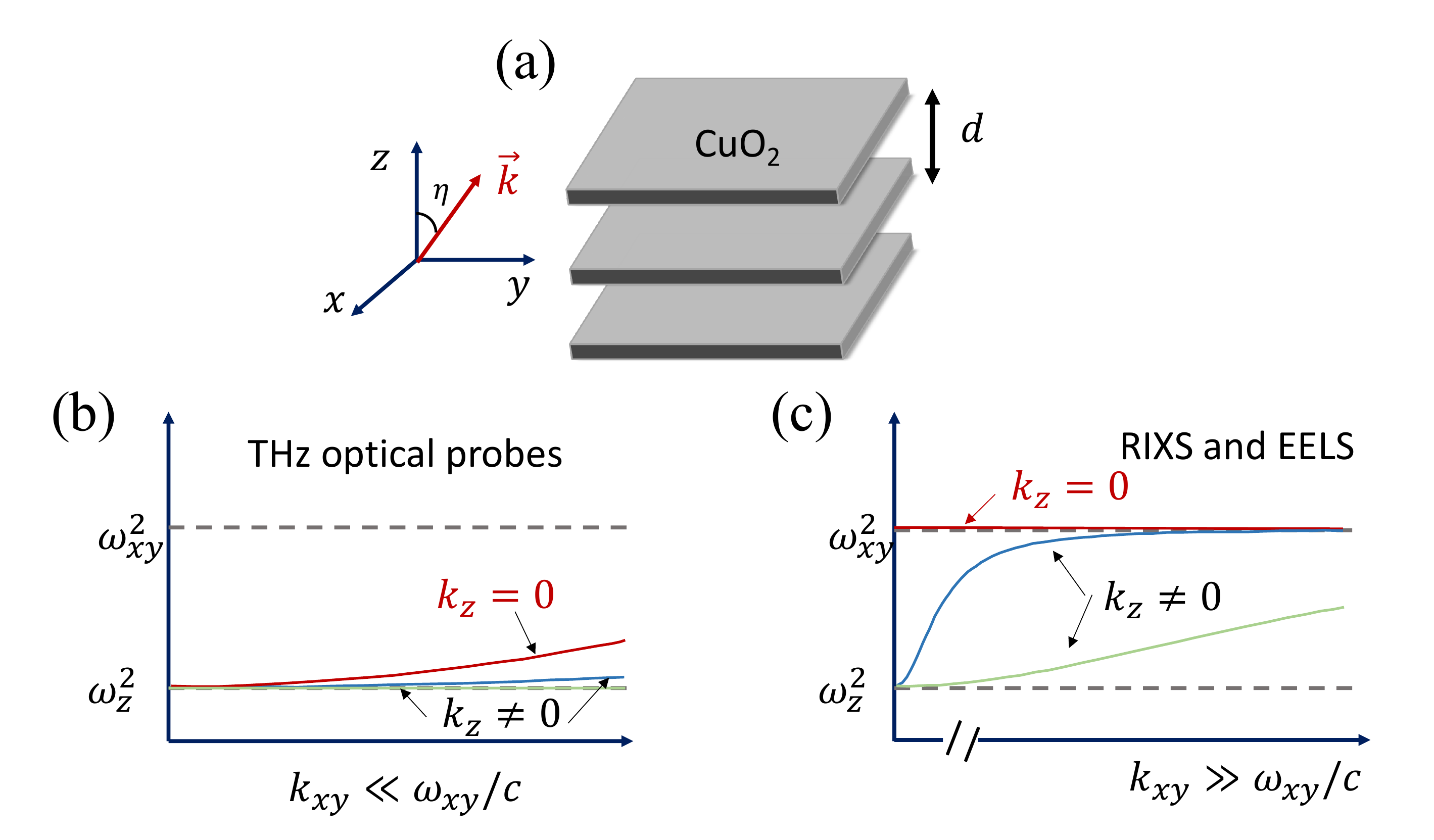}
    \caption{(a) Basic notation used in the manuscript to describe a layered superconductor. SC sheets are parallel to the $xy$ plane and stacked along $z$, at distance $d$. Within the context of cuprates the SC planes correspond to the crystallographic CuO$_2$ planes. The angle that the wavevector forms with the $z$ axis will be denoted with $\eta$. (b)-(c) Plasmon dispersion for a layered superconductor discussed so far in the literature. Panel (b) shows the dispersion \pref{small}, that is expected to describe the plasma dispersion near the low-energy out-of-plane gap $\o_z$ and for momenta much smaller than the scale set by the light cone for the largest plasma frequency. It has been derived\cite{machida_prl99,machida_physc00,bulaevskii_prb02,nori_review10,cavalleri_review,nori_natphys06} within several papers focusing on the dynamics of the out-of-plane SC phase, in relation to the THz optical response of the {Josephson plasmon}. Panel (c) shows instead the dispersion \pref{standard}, that is expected to describe the plasma dispersion near the high-energy in-plane gap $\o_{xy}$.  It is has been derived by studying the generic in-plane and out-of-plane phase or density dynamics in the SC phase\cite{vandermarel96,kresin_prb03,randeria_prb00,benfatto_prb01,millis_prr20}, and it corresponds to the standard result for the layered electron gas\cite{fetter_ap74,kresin_prb88,markiewicz_prb08,greco_cp19}, used to interpret RIXS or EELS experiments, which probes relatively high momenta and energy. }
    \lb{fig-scheme}
\end{figure}

From the experimental point view there is a strong interest in materials hosting stacks of SC sheets with weak inter-layer SC Josephson coupling\cite{nori_review10,cavalleri_review}. For the sake of concreteness, we will model them as $xy$ SC planes stacked along the $z$ direction, see Fig.\ \ref{fig-scheme}a. The benchmark example of such system is provided by high-temperature cuprate superconductors\cite{keimer_review15}. The direct consequence of the weak interlayer coupling is that the plasma energy $\o_z$  associated with phase (or density) fluctuations among the planes is substantially smaller than the one $\o_{xy}$ connected with in-plane fluctuations. In the SC state an indirect evidence of this is the large anisotropy of the in-plane vs out-of-plane penetration depths\cite{shibauchi_prl94,panagopoulos_prb96,bonn_prl04}, given by $\l_{xy/z}=c/\o_{xy/z}$. 
 Since the SC gap opening below $T_c$ suppresses  particle-hole excitations, the low-energy out-of-plane Josephson plasmon becomes undamped, and it has been observed long ago via reflectivity measurements as a clear plasma edge emerging below $T_c$ at few THz in several cuprates \cite{uchida_prl92,homes_prl93,kim_physicac95,basov_prb94,vandermarel96}. At finite momentum the dispersion of the Josephson plasmon has been derived by linearizing the sine-Gordon equations for the phase difference among the planes\cite{bulaevskii_prb94,bulaevskii_prb02,machida_prl99,machida_physc00,nori_review10,cavalleri_review,nori_natphys06}, leading to:
\be
\lb{small}
\o^2_p(\bk)=\o_z^2\left(1+\frac{c^2 \g^2 k_{xy}^2}{\o_{xy}^2+c^2  k_z^2}\right)
\ee
where $c$ is the light velocity, $\g=\o_{xy}/\o_z$ denotes the plasma-frequency anisotropy, that can be up to two orders of magnitude  in some families of cuprates, and  $k_{xy}, k_z$ denote the in-plane and out-of-plane momentum, respectively. Eq.\ \pref{small} is sketched in Fig.\ \ref{fig-scheme}b as a function of $k_{xy}$ at various values of $k_z$. As it is evident from Eq.\ \pref{small}, it represents a regular function of $\bk$ which tends always to $\o_z$ as $\bk\ra 0$. The undamped nature of this propagating mode in the SC state motivated recently several proposals of possible applications. From one side, the inherent non-linearity of the Josephson coupling among CuO$_2$ planes may enable non-linear photonic applications based on the the use of intense THz laser pulses,  as it has been widely addressed both experimentally\cite{cavalleri_review,cavalleri_natphys16,cavalleri_science18,averitt_pnas19} and theoretically \cite{nori_review10,nori_natphys06,demler_prb20,gabriele_natcomm21}. From the other side, low-energy Josephson plasmons can be attractive for photonic applications based on near-field nano-optics\cite{koppens-review-polaritons,basov-review-polaritons,basov_prb14,gozar_nqm21}, that also focus on the regime of few THz and small $k\ll \o_{xy}/c$ momenta.  

While optics is interested in the limit of long wavelength, other spectroscopic probes like resonant X-ray scattering (RIXS)  and electron energy-loss spectroscopy (EELS) are mostly able to detect the plasma dispersion around $\o_{xy}$ and relatively larger momenta $k\gg \o_{xy}/c$ as compared to the scale set by the light cone\cite{deabajo_review10}. In this regime the effects of anisotropy have been mainly discussed focusing on the emergence of acoustic-like branches of plasma excitations dispersing below $\o_{xy}$, see Fig.\ \ref{fig-scheme}c. In this case, the theoretical calculations focusing on the SC state derived the plasma dispersion either looking at the dynamics of the (in-plane and out-of-plane) SC phase\cite{randeria_prb00,benfatto_prb01,benfatto_prb04,millis_prr20} or at the dynamics of the density\cite{vandermarel96,kresin_prb03}. In both cases the standard result is fully analogous to the one found for the metallic state by means of the layered electronic model in the presence of Coulomb interactionsl\cite{fetter_ap74,kresin_prb88,markiewicz_prb08,greco_cp19}, and it reduces to a plasma dispersion given by:
\be
\lb{standard}
\o^2_p(\bk)=\o^2_{xy}\frac{k_{xy}^2}{k^2}+\o^2_z
\frac{k_z^2}{k^2}
\ee
where an additional $\sim v^2_P k^2$ term, relevant to describe the plasmon dispersion with velocity $v_P$ at $k_z=0$, can be added\cite{millis_prr20}. As one can see in Fig.\ \ref{fig-scheme}c, at finite $k_z$ there is an intermediate regime of $k_{xy}$ values where the plasmon softens below $\o_{xy}$ crossing from a quasi-linear to a $\o\propto \sqrt{k}$ behavior, typical of two-dimensional plasmons.  So far, such acoustic-like branches have been identified by RIXS in some high-temperature cuprates\cite{lee_science18,liu_nqm20,zhou_prl20}, while EELS 
observed only the high-energy optical branch, with a strong and still not well understood overdamping\cite{mitrano_pnas18,mitrano_prx19,fink_cm21,mitrano_cm21}. Despite the fairly good agreement between Eq.\ \pref{standard} and RIXS experiments, which probe energies around $\o_{xy}$ and large momenta as compared to $\o_{xy}/c$, there  is apparently no connection between Eq.\ \pref{standard} and the expression \pref{small}, which should represents the limit of the plasma dispersion when one focuses around $\o_{z}$ and small momenta. Indeed, in contrast to Eq.\ \pref{small}, Eq.\ \pref{standard} is non-analytic as $\bk\ra 0$, since it predicts a continuum of possible values which depend on the angle that the wavevector $\bk$ forms with $z$ axis. In addition, at $k_z=0$ and finite $k_{xy}$, as shown in Fig.\ \ref{fig-scheme}b,c, Eq.\ \pref{small} predicts a mode dispersing away from the soft Josephson plasmon $\o_z$ while Eq.\ \pref{standard} (with the additional $v_P^2 k^2$ term) predicts a mode around $\o_{xy}$. 

In this paper we provide a general derivation of the plasma dispersion in a layered superconductor which is valid at generic momentum, and we identify a crossover value $k_c=\sqrt{(\o_{xy}^2-\o_z^2)}/c$ such that the two previously reported expressions \pref{small} and \pref{standard} emerge as the limit of small $k\ll k_c$ or large $k\gg k_c$ momentum of the full dispersion. To this aim we analyze the problem focusing on the degrees of freedom linked to the SC phase of the order parameter within an effective-action formalism, which naturally implements the spontaneous symmetry breaking below $T_c$. We introduce explicitly the electromagnetic (e.m.) scalar and vector potentials to construct a gauge-invariant quantum action for the coupled system given by the matter (represented by the SC phase) and the e.m. fields. The main technical difference with respect to previous work using the same formalism\cite{randeria_prb00,benfatto_prb01,millis_prr20}, that was able to obtain only the result Eq.\ \pref{standard}, relies on the fact that for an anisotropic superconductor the role of e.m. interactions is not exhausted by the Coulomb interaction, which originates from the scalar potential. The reason is that, as one already observes at the level of Maxwell equations\cite{bulaevskii_prb94,bulaevskii_prb02,machida_prl99,machida_physc00,demler_prb20}, in a layered system the e.m. longitudinal and transverse response get intrinsically mixed. The physical reason is that once the superfluid response is anisotropic the induced current is no more parallel to the electric field. Within the context of the SC phase dynamics the anisotropy of the superfluid response leads, in contrast to the isotropic case, to a finite coupling of phase fluctuations to both the longitudinal and transverse components of the e.m. fields.  When both couplings are properly included the phase spectrum at arbitrary wave vector carries on the information on a mixed longitudinal-transverse excitation, except for the two special limits  $(k_{xy}\ra0,k_z=0)$ or $(k_{xy}=0,k_z\ra0)$. By introducing an appropriate gauge-invariant phase variables we can derive an analytical expression for the generalized plasma modes valid at generic momentum. In particular, the mode lower in energy, that becomes purely longitudinal only well above $k_c$,  if shown to interpolate among the two limiting behaviors provided by Eq.\ \pref{small} and \pref{standard}, clarifying then the limits of their validity. In addition, such a formulation allows one to easily extend the result to the case of a non-linear Josephson model,  as previously discussed within the context of the out-of-plane Josephson plasmon\cite{bulaevskii_prb94,bulaevskii_prb02,machida_prl99,machida_physc00,demler_prb20}. Finally, in order to make a closer connection with experimental probes sensitive to density correlations\cite{deabajo_review10}, we explicitly derive the density-density response in the general case, showing that both generalized plasma modes appear in the spectrum of the physical observables, with an equal weight at the length scale $\sim 1\k_c$, which is typically of the order of a fraction of micron. At larger momenta $k\gg k_c$ the density response closely matches the standard RPA result, and a single quasi-longitudinal plasma mode is visible in the density spectral function. Our results represent a complete  gauge-invariant description of generalized plasma excitations in a layered superconductor valid at different energy and momentum scales, providing the benchmark behavior to analyze plasmonic effects in a wide set of experiments ranging from near-field optics to RIXS and EELS.

The plan of the paper is the following. In Sec. \ref{isotropic} we review the standard derivation of the Gaussian phase-only action in the isotropic case. The Section starts with an introductory subsection where we outline the theoretical approach used in the manuscript, and a second one where we show its application in the standard isotropic case. Here we show that the plasma mode appears in the different sectors of the matter-e.m. field action depending on the gauge choice, but leading always to the same result.  The results of this Section are not new, but their derivation is carried out in a different way with respect to previous work, and it allows us to introduce the formalism relevant for the rest of the paper. In Sec. \ref{layered} we derive the action for the layered 3D case, expressed in terms of the appropriate gauge-invariant variables. We then show that the longitudinal and transverse e.m. excitations are always coupled, and both excitations appear in the phase spectrum. A simplified description of the problem within the context of Maxwell's equation is also provided in Appendix B. In subsection \ref{sinegordon} we rephrase the same results in terms of approximate equations of motions, that has been widely used in the recent literature\cite{bulaevskii_prb94,bulaevskii_prb02,machida_prl99,machida_physc00,nori_review10,cavalleri_review,nori_natphys06} in order to include non-linear effects of the plasmon dynamics. In Sec. \ref{Density}, we show how the mixed longitudinal-transverse modes found in Sec. \ref{layered} also appears as poles of the properly computed density-density response function. Sec. \ref{Concl} contains a general discussion about the results. Further technical details are provided in the Appendixes. Appendix A reviews the main steps leading to the effective action for a superconductor within the functional-integral formalism. Appendix B shows the derivation of the mixed longitudinal-transverse modes from the solution of the Maxwell's equations.  


\section{Effective action for the isotropic case}\label{isotropic}

\subsection{Description of the plasmon via the SC phase}

Before giving technical details about the derivation of the plasma dispersion for the layered superconductor, it can be useful to briefly outline the general context, starting from the isotropic case. As mentioned in the introduction, the SC transition carries on specific signatures both in the single-particle fermionic excitations spectrum, via the opening of the SC gap below $T_c$, and in the spectrum of the bosonic-like collective excitations. More specifically, while the gap is connected to the equilibrium value of the SC complex order parameter, two modes emerge connected to fluctuations of its amplitude and phase. Since the SC order parameter breaks the continuous gauge symmetry  a Goldstone mode is expected, that is directly linked to phase fluctuations of the SC order parameter\cite{nagaosa}. There are several ways to see this. At the level of the classical Ginzburg-Landau description this appears as an energetic cost needed to "twist" the phase $\th$  with respect to its equilibrium value. In full analogy with the elastic deformation of a solid body the cost $E$ of a finite phase gradient {\em in space} scales with a superfluid "stiffness" $D_s$, such that 
\be
\lb{classical}
E=\frac{\hbar^2 D_s}{8}\int d\bx \left(\bdnb \th\right)^2.
\ee
In the simplest Galilean-invariant case $D_s$ can be  expressed as the ratio between the density of superfluid electrons and their mass, $D_s=n_s/m$, with $n_s=n$ at $T=0$.  As the length scale of spatial  phase deformations goes to zero $E$ vanishes. This can be interpreted as a mode at zero energy for $\bk=0$, that is what expected for a Goldstone mode. On the other hand, to really define a propagating mode, i.e. to establish a frequency-momentum dispersion, one needs to include dynamical effects, not present in the classical Ginzburg-Landau description. A very elegant and powerful technique relies\cite{nagaosa} on the explicit construction of a quantum analogous of the Ginzburg-Landau model by starting from a microscopic interacting  model for electrons. The basic idea is to start from a fermionic model with a BCS-like interaction term and decouple it via the Hubbard-Stratonovich procedure by introducing two effective bosonic fields which play the role of the order-parameter amplitude and phase. By further integrating out explicitly the fermions one obtains a quantum model expressed in terms of the collective SC variables, that can be expanded in principle up to arbitrary powers in the collective-mode fields, linking the phenomenological couplings of the Ginzburg-Landau model  to fermionic susceptibilities, which contain all the information on the microscopic physics under investigation.
Such a procedure is outlined  in Appendix A, and further details can be found in several books and papers on the subject, see e.g. Ref.s\ \cite{nagaosa,aitchison_prb95,depalo_prb99,randeria_prb00,benfatto_prb01,benfatto_prb04,millis_prr20}. By retaining only Gaussian terms in the fluctuations one defines the spectrum of the collective modes, equivalent to compute the response functions at RPA level in the usual diagrammatic approach to fermionic models\cite{nagaosa,pines}. 
Within this quantum generalization of the classical Ginzburg-Landau theory one finds that the density appears as conjugate variable of the phase\cite{aitchison_prb95,depalo_prb99,nagaosa}. As a consequence, within the quantum phase-only action the energetic cost to perform a phase gradient {\em in time } is controlled by the charge compressibility $\kappa_0$\cite{aitchison_prb95,depalo_prb99}, and Eq.\ \pref{classical} is replaced at $T=0$ by the quantum action: 
\be
\lb{sound}
S=\frac{\hbar^2}{8}\int dt d\bx \left[ \k_0 \left(\pd_t \th\right)^2-D_s \left(\bdnb \th\right)^2\right].
\ee
For weakly-interacting neutral systems $\k_0$ in the static long-wavelength limit can be approximated with the density of states at the Fermi level, and in Eq.\ \pref{sound}  one recognizes the so-called\cite{anderson_pr58} Anderson-Bogoliubov sound mode:
\be
\lb{neutral}
\o^2=v_s^2 |\bk|^2
\ee
where $v_s^2=D_s/\k_0$ is the sound velocity.

The identification of the density as conjugate variable of the phase is a general result,  which holds both for neutral superfluids and for superconductors, and has profound implications for the latter. Indeed, in the standard description of charged systems, density-density interactions are expected to be mediated by the long-range Coulomb potential. In the literature based on the effective-action formalism the difference between neutral and charged superfluids is then usually encoded\cite{nagaosa,depalo_prb99,randeria_prb00,benfatto_prb01,benfatto_prb04,millis_prr20} via an additional Hubbard-Stratonovich field which decouples density-density fermionic interactions, so that its integration 
dresses the charge compressibility at RPA level by the Coulomb potential\cite{anderson_pr58}. The term $\k_0$ in Eq.\ \pref{sound} is then replaced in Fourier space  by $\k(\bk)=\k_0/(1+V(\bk)\k_0)$, where $V(\bk)$ is the Coulomb potential in generic D dimensions. Since for $\bk\ra 0$ one has $\k\ra 1/V(\bk)$ the spectrum of the phase mode, that reflects the one of density fluctuations, identifies now a plasma mode, whose energy vs momentum dispersion depends on the dimensions. In the standard isotropic three-dimensional (3D) case one recovers the well-known dispersion
\be
\lb{omegap}
\o^2=\o_p^2+v_s^2 |\bk|^2
\ee
where $\o_p^2=4\pi e^2 D_s$ is the isotropic plasma frequency.

The procedure outlined above explains why in the SC state the spectrum of the collective plasma excitations can be derived in a relatively simple way by analyzing the SC phase fluctuations. However, this analysis is not enough if one wants to determine the spectral weight of the plasma mode, that can only be determined by direct derivation of the density-density response\cite{nagaosa,randeria_prb00,benfatto_prb04,millis_prr20}, which couples to external probes. Indeed, on a more general ground, the SC  phase itself does not represent a physically observable quantity. Once the matter is coupled to the e.m. field, the physical observables are related to the gauge-invariant four-vector $\psi_\m$, whose temporal and spatial components read:
\be
\lb{giphase}
\psi_0 \equiv 
\pd_t \th+\frac{2e}{\hbar }\phi \quad
\bpsi \equiv \bdnb \th - \frac{2e}{\hbar c}\bA
\ee
where $\phi$ and $\bA$ are the scalar and vector potentials respectively. In contrast to the phase alone, the observables \pref{giphase} are invariant under the simultaneous gauge transformation of the e.m. potentials (that leaves the physical electrical and magnetic field unchanged) and of the SC phase:
%
%
\be
\lb{gautran}
\begin{cases}
&\th(\bx,t) \ra \th(\bx,t)+\frac{2e}{\hbar c}\l(\bx,t),
\\
\\
& \bA (\bx,t) \ra \bA (\bx,t)+\nabla  \l(\bx,t),
\\
\\
& \phi (\bx,t) \ra \phi(\bx,t) -\pd_t  \l(\bx,t).
\end{cases}
\ee
In this picture one  realizes that analyzing the spectrum of the phase mode in the superconductor is completely equivalent to solve the problem of the e.m. wave propagation. More specifically, the frequency-momentum relation \pref{omegap} corresponds to the one of the longitudinal component of the electric field, which couples by Maxwell's equations to charge fluctuations and then  to phase fluctuations in a superconductor. In general, e.m. waves can also be transverse, but in the isotropic system they are not coupled to phase fluctuations. Once again, this result can be easily understood already at classical level. In isotropic and homogeneous systems the minimal-coupling substitution \pref{giphase} into the gradient term \pref{classical} leads to 
\be
\lb{classical2}
E=\frac{\hbar^2 D_s}{8}\int d\bx \left(\bdnb \th -\frac{2e}{\hbar c}\bA\right)^2.
\ee
The explicit coupling among the phase and the gauge field in Eq.\ \pref{sound} is the term $\int d\bx D_s(\bdnb \th)\cdot \bA$, that for a constant $D_s$ can be written after integrating by part as $\int d\bx \, D_s \th (\bdnb\cdot \bA)$. As a consequence the phase  couples {\em only} to the longitudinal component of the e.m. field, and one does not see any signature of the transverse e.m. mode in the phase spectrum. 
Such a decoupling of the longitudinal and transverse sectors makes the description of the plasma mode completely equivalent in the two approaches: via the matter, where the charged nature of the system enters via the Coulomb interaction for the density, or via e.m. potentials, where one derives directly the equations of motions for the  e.m. fields. 

The situation becomes radically different for {\em anisotropic} systems, and in particular for layered superconductors. In this case, which is relevant for several classes of unconventional superconductors, as e.g. cuprates or pnictides, the phase stiffness in the plane $D_\pll$ and in the direction perpendicular to the planes $D_\perp$ is anisotropic, with $D_\perp\ll D_\pll$\cite{shibauchi_prl94,panagopoulos_prb96,bonn_prl04}. Within the language of Eq.\ \pref{classical2} above, this implies that the phase can also be coupled to the transverse vector potential. This is the central physical effect which motivates the present work, aimed to explain what are the consequences of such a mixing of the transverse and longitudinal modes for the spectrum of plasma waves. Indeed, as we mentioned in the introduction, so far the effect of this mixing has been included in an approximated way in the work\cite{machida_prl99,machida_physc00,bulaevskii_prb02} aimed to derive the dispersion \pref{small} of the soft Josephson plasmon, but it is completely missing in any approach extending the RPA to the layered electron gas\cite{fetter_ap74,kresin_prb88,markiewicz_prb08,greco_cp19} or to the layered superconductors\cite{randeria_prb00,benfatto_prb01,millis_prr20,vandermarel96,kresin_prb03}. Indeed, in the latter case, which leads to the expression \pref{standard} for the layered plasmon, one  takes into account only the effect of the Coulomb potential, which is linked to the longitudinal component of the e.m. field. 

In Sec.\ \pref{layered} we will explicitly show how the transverse-longitudinal mixing appears in the quantum phase-only model for the layered superconductor, and how one can derive from it an analytical expression for the generalized plasma dispersion which interpolates between the two limits \pref{small} and \pref{standard}, clarifying at the same time their range of validity. To get a better insight into the link among the phase mode and the plasma wave we will first review in the next subsection the isotropic case. In contrast to previous work\cite{nagaosa,depalo_prb99,randeria_prb00,benfatto_prb01,millis_prr20} where the e.m. field only enters via the Coulomb density-density interaction, we introduce here the coupling \pref{giphase} of the SC phase to the internal e.m. field, that we further integrate out. We then show how the same plasmon dispersion is obtained regardless the gauge choice for the e.m. field, as expected. This approach will then be extended to the layered case in Sec.\ \pref{layered}.

\subsection{Formulation of the problem with the internal e.m/ fields}
As a starting point we use the imaginary-time equivalent of the quantum action \pref{sound} for an isotropic 3D superconductor, rewritten in Fourier space. As mentioned above, its derivation relies on a rather standard technique\cite{nagaosa,aitchison_prb95,depalo_prb99,randeria_prb00,benfatto_prb01,benfatto_prb04,millis_prr20}, whose details are given in Appendix A. We then have:
\be
\lb{bare}
S_G[\th]=\frac{1}{8}\sum_{q}
\left[
\k_0\O_m^2+D_s|\bk|^2
\right] |\th(q)|^2 
\ee
where $q\equiv(i\O_m,\bk)$ is the imaginary-time 4-momentum, $\O_m=2\pi m T$ are bosonic Matsubara frequencies, $\k_0$ is the bare compressibility, and $D_s$ is the superfluid stiffness. In the following we put $\hbar=k_B=1$. In ordinary Maxwell equations the source for the e.m. fields are provided both by external charge/currents and by the induced internal charge/current fluctuations in the matter. In the superconductor the latter will be described by the SC phase, coupled to the internal e.m. fields via the minimal coupling \pref{giphase}. To account for the e.m. fields we then need to include first the free contribution, which reads\cite{nagaosa}:
\bea
&&S_{e.m.}[A_{\m}]=\int d\t d\bx \left[\frac{(\bdnb\times\bA)^2}{8\pi}+\right.\nn\\
-\frac{\e}{8\pi}& &\left.\left(\frac{i\pd_\t \bA}{c}+\bdnb \phi\right)^2\right]=\frac{\e}{8\pi}\sum_q \left[ \frac{\O^2_m}{c^2} |\bA(q)|^2+\right.\nn\\
&&-|\bk|^2 |\phi(q)|^2+\frac{|\bk|^2}{\e} |\bA_T(q)|^2+\nn\\
\lb{sem}
+&&\frac{i\O_m}{c}\left.\bk\cdot \left( \phi(q)\bA_L(-q)+\phi(-q)\bA_L(q)\right)\right]
%
\eea
Here we introduced the longitudinal $\bA_L=(\hat \bk \cdot \bA)\hat \bk$ and transverse $\bA_T=\bA-\bA_L=(\hat \bk\times \bA)\times\hat{\bk}$ components of $\bA$.  Eq.\ \pref{sem} is just a transcription of the usual form $\frac{-\mid \mathbf{E} \mid^2+\mid\mathbf{B} \mid^2}{8\pi}$, where the electric field is expressed in the imaginary-time formalism as $\mathbf{E}=-\frac{i}{c}\pd_\t \bA-\bdnb \phi$. It is worth noting that in order to have a definition of $\bE$ analogous to the one valid for real time one should assume that $\phi$ is purely imaginary, i.e. one should replace $\phi\ra i\phi$. In this case, by defining the imaginary-time electric field as $\bE\equiv-\frac{1}{c}\partial_\t\bA-\bdnb\phi$ the action for the free e.m. field would read $\frac{\e|\bE|^2+|\bB|^2}{8\pi}$, which is the usual expression for the energy density. Such a rescaling of the scalar potential would also make the quadratic term in the scalar potential arising from $\left(\bdnb\phi\right)^2$ positive defined, as required to perform the Gaussian integration. To make notation more compact we will not explicitly rescale the potential in what follows, but we will implicitly assume that a formal definition of the Gaussian integration in the imaginary-time formalism requires such a regularization. In order to include the ionic screening, we also introduced the background dielectric constant $\e$. As mentioned above, in the superconductor\cite{nagaosa,depalo_prb99,randeria_prb00,benfatto_prb04} the coupling of the e.m radiation with the matter can be easily implemented by using the SC phase, via the minimal coupling substitution \pref{giphase}, which reads in momentum space and Matsubara frequency:
\be
\lb{mincoup}
\begin{cases}
& \O_m \th(q)\ra \O_m \th(q)+2e \phi(q), \\
\\
& i\bk \th(q) \ra i\bk \th(q)-\frac{2e}{c}
\bA(q)\\
\end{cases}
\ee
%
 By replacing Eqs.\ \pref{mincoup} into the action \pref{bare} one obtains a light-matter coupling term $S_{\th A_{\m}}[\th,A_{\m}]$ and a renormalization of the bare e.m. action, so the total action reads:
\be
\lb{TotalAction}
S[\th,A_{\m}]=S_G[\th]+S'_{e.m.}[A_{\m}]+S_{\th A_{\m}}[\th,A_{\m}]
\ee
where $S'_{e.m.}[A_{\m}]$ describes the long-wavelength propagation of light through matter. 
For the isotropic system this term reads:
\bea
\lb{semtot}
S'_{e.m.}[A_{\m}]&&=\frac{\e}{8\pi}\sum_q
\left[-\left(|\bk|^2+k^2_{TF}\right)|\phi(q)|^2+\right.\nn\\ 
&&+\left(
\O_m^2+\o_p^2
\right)\frac{|\bA_L(q)|^2}{c^2}+\nn\\
&&+
\left(\O_m+\frac{c^2}{\e}|\bk|^2+ \o_p^2 \right)
\frac{|\bA_T(q)|^2}{c^2} +\nn\\
+\frac{i\O_m}{c}&&\left.\bk\cdot\left(\phi(q)\bA_L(-q)+\phi(-q)\bA_L(q)\right)\right].
\eea
In the effective-action formalism the frequency-momentum dispersion obtained as solution of the Maxwell's equations appears as the zero of the determinant of the matrix associated with the coefficients of the fields in the Guassian action, once the analytical continuation $i\O_n\ra \o+i\d$ is performed. As a consequence, one clearly sees that the inclusion of the matter has two well-known effects on the e.m. fields: (i) the scalar potential displays the usual Thomas-Fermi screening, where $k^2_{TF}=4\pi e^2 \k_0/\e$; (ii) the 
propagating transverse mode
\be
 \lb{transv}
 \o^2_T=\o_p^2+\ct^2|\bk|^2
 \ee
is gapped out\cite{anderson_pr63} at the isotropic (screened) SC plasma frequency%
\be
\lb{omp}
\o_p^2\equiv 4\pi e^2 D_s/\e,
\ee
and the light velocity is renormalized as $\ct=c/\sqrt{\e}$. In the typical field-theory language, the Anderson-Higgs mechanism\cite{anderson_pr63} manifest as a "mass" term $\o_p^2 \bA_T^2$ in the e.m. action \pref{semtot}, that, in the static limit, only survives below $T_c$, where $D_s\neq 0$. 
Finally, the coupling between the phase and the e.m. potential reads:
\bea
 S_{\th A_{\m}}&&[\th,A_{\m}]=\frac{e}{4} \sum_q
\left[ \k_0 \O_m \left(\th(q)\phi(-q)-\th(-q)\phi(q)\right)+\right.\nn\\
\lb{scoup}
&&\left.-i\frac{D_s}{c}\bk\cdot\left(\th(q)\bA_L(-q)-\th(-q)\bA_L(q)\right) \right].
\eea
The last term of Eq.\ \pref{scoup} shows explicitly that in the isotropic  case phase fluctuations only couple to the longitudinal component of the vector potential $\bA_L$, as we mentioned above. In addition, the use of Eq.\ \pref{mincoup} clearly guarantees that the total action is invariant under the gauge transformation \pref{gautran}, which reads explicitly in Matsubara formalism:
\be
\begin{cases}
&\th(q)\ra\th(q)+\frac{2e}{c}\l(q)
\\
\\
& \phi(q) \ra \phi(q) -\frac{\O_m}{c}\l(q)
\\
\\
& \bA(q)\ra\bA(q)+i\bk \l(q)
\end{cases}
\label{GaugeInvariance}
\ee
$\l(q)$ being an arbitrary function of the Fourier momenta.  Such a gauge freedom also implies that only the gauge-invariant combination \pref{mincoup} represents a physically-observable quantity describing matter properties, while the information carried out by the SC phase alone acquires a different meaning within different gauge choices. This means that even though the phase-only propagator is always linked to the plasma mode, it will appear in different forms depending on the gauge. 

To clarify this point, let us start from Eq.\ \pref{TotalAction} and let us derive the action for the phase degrees of freedom by integrating out the e.m. potentials. A first natural gauge choice is the so-called Coulomb gauge $\bdnb\cdot\bA=0$, i.e. $\bA_L=0$. Indeed, as noticed below Eq.\ \pref{scoup}, in the isotropic case the phase fluctuations only couple to  $\bA_L$, so that in the Coulomb gauge only the coupling between the phase and the scalar potential survives. By integrating out $\phi$ one is then left with:

\be
\lb{isocg}
S_{\bdnb\cdot\bA=0}^{(iso)}
=
\frac{\e}{32\pi e^2}\sum_q \left[\frac{\O_m^2}{1+\a|\bk|^2}+
\o_p^2
\right]
|\bk|^2
|\th(q)|^2
\ee
where we defined $\a$ as
\be
\lb{defa}
 \a\equiv \frac{\e}{({4\pi e^2 \k_0})}=\frac{1}{k_{TF}^2}=\l_D^2
 \ee
where $\l_D$ is the Debye screening length. The result \pref{isocg} is  formally identical to the one widely discussed in the previous literature\cite{nagaosa,depalo_prb99,randeria_prb00,benfatto_prb04,millis_prr20}, and obtained by adding a Coulomb-mediated density-density interaction term in the starting microscopic fermionic model. This is a natural consequence of the fact that in the Coulomb gauge only the coupling of the phase to the scalar potential is relevant.

An alternative but completely equivalent approach can be followed integrating out the e.m. fields in the Weyl gauge, where  $\phi=0$. In this case only the coupling to $\bA_L$ survives in Eq.\ \pref{scoup} and one obtains:
\be
\lb{isowg}
S_{\phi=0}^{(iso)}
=
\frac{\k_0}{8}\sum_q
\frac{\O_m^2}{\O_m^2+\o_p^2}\left[\O_m^2+
\left(1+\a|\bk|^2\right)
\o_p^2\right] |\th(q)|^2
\ee
We immediately see that both \pref{isocg} and \pref{isowg} identify a collective excitation as a pole of the $\th(q)$ fluctuations, which are controlled by the inverse of the $|\th(q)|^2$ term. After analytical continuations to real frequencies this is given by:
\be
\lb{omdisp}
\o^2=\o_p^2(1+\a |\bk|^2)
\ee
that is the usual dispersion of the longitudinal plasma mode in a superconductor. 
As usual, the plasmon velocity $c_p=\sqrt{\a}\o_p=\ct (\l_D/\l)$, where $\l$ is the London penetration depth, is much smaller than the light velocity $\ct$ of the transverse wave in the medium, since $\l_D/\l\ll 1$ even for unconventional superconductors. 
It is worth noting that the velocity of the plasmon in Eq.\ \pref{omdisp} is not the same as the one obtained for the normal metal\cite{fetter}, that would correspond to $\o^2=\o_p^2(1+(9/5)\a|\bk|^2)$. This is due to the fact that to correctly account for the plasma dispersion one should also account for the $|\bk|^2$ corrections to the BCS density-density response function in the SC state,  that appears as a coefficient of the $(\pd_t\th)^2$ term in the phase-only action, while in Eq.\ \pref{sound} only its long wave-length limit $\kappa_0$ has been retained. To simplify the notation we shall not perform explicitly such an expansion, that is anyway irrelevant to the physical effects we want to discuss in the present manuscript. Indeed, it only gives a small quantitative difference that can be included if one is interested into a detailed quantitative comparison with the experiments.

While in Eq.\ \pref{isocg} the plasma mode appears in the spectrum of $\bdnb \th$ fluctuations, in the Weyl gauge it appears in the spectrum of $\pd_\t \th$ fluctuations. This is a direct consequence of the fact that the phase variable does not represent a true physical observable. In addition, it describes, as expected,
only the longitudinal mode. For both gauge choices the transverse e.m. mode is described by the  two remaining degrees of freedom of the e.m. field in Eq.\ \pref{TotalAction}. A very elegant and convenient  way to derive simultaneously the energy-momentum dispersion for all the transverse and longitudinal e.m. modes relies on the use of the gauge-invariant physical observables \pref{giphase}. For example one can use the spatial component of the gauge-invariant (g.i.) phase difference:
\be
\lb{gipd}
 \bpsi=\bdnb\th-\frac{2e}{c}\bA
\ee
and set to zero the scalar component via a proper gauge fixing. The equivalent of Eq.\ \pref{TotalAction} in the new variables reads:
%
\bea
\lb{sthpsi}
&&S[\th,\bpsi]=\frac{\e}{32\pi e^2}\sum_q \left[\left(\O_m^2+\o_p^2\right) |\bpsi(q)|^2+\right.\nn\\\
&+&\left.\ct^2\left|\bk\times\bpsi\right|^2+\frac{\O_m^2}{\a}\left(
1+\a|\bk|^2\right)|\th(q)|^2+\right.\nn\\\
&&+\left.i\O_m^2 \bk\cdot \left(\bpsi(q) \th(-q)-\bpsi(-q) \th(q)\right)\right]
\eea
By introducing again the longitudinal $\bpsi_L=(\hat\bk\cdot \bpsi)\hat\bk$ and the transverse $\bpsi_T=\bpsi-\bpsi_L=(\hat{\bk}\times\bpsi)\times\hat{\bk}$ components we see that only $\bpsi_L$ couples to the phase $\th$.  After integrating it out one then finds:
\bea
 S[\bpsi]&=&\frac{\e}{32\pi e^2}\sum_q\left[
\left(\frac{\O_m^2}{1+\a\abs{\bk}^2}+\o_p^2\right)|\bpsi_L(q) |^2+\right.\nn\\
\lb{isopsi}
&+&\left.\left(\O_m^2+\o_p^2+\ct^2 \abs{\bk}^2\right)|\bpsi_T(q)|^2
\right]
\eea
From Eq.\ \pref{isopsi} one immediately sees that the three components of  $\bpsi$ describes all e.m. modes, as given by the poles of the longitudinal and transverse propagators. Indeed,  after analytical continuation, one finds the solution \pref{omdisp} in the longitudinal sector and the solution \pref{transv} in the transverse sector, respectively.  Such a result suggests that in the anisotropic case where longitudinal and transverse modes get mixed a description in terms of the physical fields \pref{gipd} can be more convenient, as we shall see indeed in the next section. In addition, this choice will make the extension of the Gaussian action to a non-linear Josephson model straightforward, since it will only affect the mass term for the $\bpsi$ field, as we will see in details in Sec. \ref{sinegordon}.

\section{Effective action for a layered 3D system}\label{layered}
\begin{figure}[ht]
    \centering
    \includegraphics[width=0.5\textwidth,keepaspectratio]{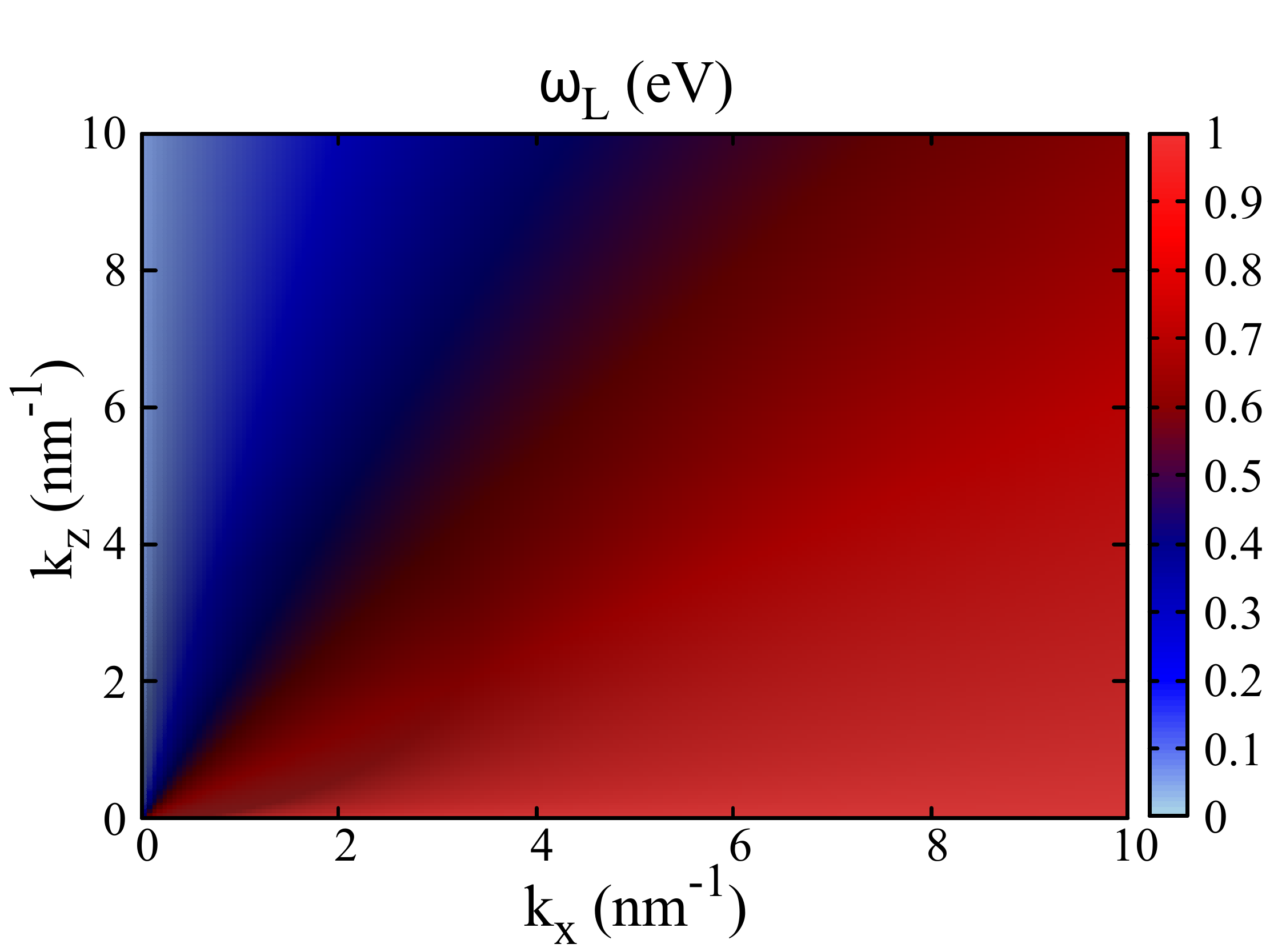}
    \caption{(a) Plasmon dispersion given by Eq.\ \pref{opmillis}, as obtained by ignoring the coupling of the SC phase to the transverse vector potential. As explained in the text, it also corresponds to the standard result obtained by computing the density-density response in RPA in an anisotropic superconductor\cite{randeria_prb00,benfatto_prb01,benfatto_prb04,millis_prr20,vandermarel96,kresin_prb03} or or metal\cite{fetter_ap74,kresin_prb88,markiewicz_prb08,greco_cp19}. Selected cuts of $\o_L$ at fixed $k_z$ as a function of $k_x$ are reported in Fig.\ \ref{fig-scheme}c. }
    \lb{fig-RPA}
\end{figure}
\subsection{Derivation of the plasma dispersion for the anisotropic system}

To discuss the layered case we should consider an extension of Eq.\ \pref{bare} where the stiffness is anisotropic. By using the notation of Fig.\ \ref{fig-scheme}a we will denote with $D_{xy}$  and $D_z$  the in-plane and out-of-plane stiffness, respectively. Eq.\ \pref{bare} can be straightforwardly generalized\cite{randeria_prb00,benfatto_prb01,millis_prr20} to:
\be
\lb{barelay}
S_G[\th]=\frac{1}{8}\sum_{q}
\left[
\k_0\O_m^2+D_{xy} k_{xy}^2+D_z k_z^2
\right]\th(q)\th(-q),
\ee
where $k_{xy}\equiv\sqrt{k_x^2+k_y^2}$ is the in-plane momentum. Notice that, despite of the discrete out-of-plane structure of the system, we still use a continuum anisotropic model, as valid when $k_z\ll 1/d$. Indeed, the breakdown of the standard RPA approximation and the mixing between longitudinal and transverse degrees of freedom, which will be the main topics this section, occur at very small momenta $\bk$, which are suitably accounted for in the continuum limit. The anisotropy of the stiffness has important consequences when the electrons are coupled to the e.m. fields via the minimal-coupling substitution \pref{mincoup}. Indeed, the action \pref{semtot} describing the e.m. field in the matter gets modified as:
\bea
\lb{semtotlay}
&&S'_{e.m.}[A_{\m}]=\frac{\e}{8\pi}\sum_q
\left[-(k_{TF}^2+|\bk|^2)|\phi(q)|^2+\right.\nn\\
&&+\frac{1}{\e}|\bk \times \bA(q)|^2+
\frac{(\O_m^2+\o_{xy}^2)}{c^2}
|\bA_{xy}(q)|^2+\nn\\
+&&
\left.\frac{(\O_m^2+\o_z^2)}{c^2}|\bA_{z}(q)|^2+
\frac{i\O_m}{c}\phi(q)\bk\cdot \bA(-q)-h.c.\right]\nn\\
\eea
where we defined the two plasma frequencies:
\be
\lb{omplay}
\o_{xy}^2=\frac{4\pi e^2 D_{xy}}{\e}, \quad \o_z^2=\frac{4\pi e^2 D_z}{\e}
\ee
In principle, in Eq.\ \pref{semtotlay} we could also consider an anisotropic background dielectric constant, in order to take into account the different ionic screening occurring along the in-plane and the out-of-plane directions. The e.m. energy density would then read $\frac{-\e_{\a}|\mathbf{E}_{\a}|^2+|\mathbf{B}|^2}{8\pi}$, with $\e_{\a}=\e_{xy}$ for $\a=x,y$, and $\e_{\a}=\e_{z}$ for $\a=z$. However, this effect does not modify substantially the physics of the system, therefore for the sake of simplicity we take $\e$ isotropic.\\ 
Finally, the coupling of the e.m. fields with the phase variable is an anisotropic version of Eq.\ \pref{scoup}, that is now expressed as: 
\bea
 S_{\th A_{\m}}[\th,A_{\m}]&=&\frac{e}{4} \sum_q
\left[ \k_0 \O_m\th(q)\phi(-q)+h.c+\right.\nn\\
&-&\frac{D_{xy}}{c}i\th(q)\bk_{xy}\cdot{\bA}_{xy}(-q)+h.c.+\nn\\
\lb{scouplay}
&-&\left.\frac{D_z}{c}i\th(q)\bk_z\cdot \bA_z(-q)+h.c.\right]
\eea
As one can see, the anisotropy of the SC phase stiffness has two main consequences in the description of the e.m. response: (i) the massive terms in the vector potential of Eq.\ \pref{semtotlay} become anisotropic; (ii) the inclusion of the matter via the SC phase shows that there is no way to decouple completely the longitudinal and transverse modes at arbitrary wave vector $\bk$. The latter result emerges more clearly by close inspection of Eq.\ \pref{scouplay}. While in the isotropic case the equivalent term $D_s \bk\cdot\bA$ of Eq.\ \pref{scoup} implies that $\th$ only couples to the longitudinal component of the vector potential, in the anisotropic case 
the combination $D_{xy}\bk_{xy}\cdot \bA_{xy}+D_z\bk_z\cdot \bA_z$ is different from zero even in the Coulomb gauge where $\bk\cdot \bA=0$. This is the crucial point that has been missed in previous work where the effect of the anisotropy has been only included in the phase stiffness, but not on the coupling to the gauge potential. At a more general level, the physical effect we are implementing here is the fact that the superfluid current is no more parallel to to electric field, so even a purely longitudinal field always induces a transverse current and viceversa. Such an interpretation is straightforward when the problem is studied via Maxwell's equations only, as we discuss in Appendix B.

To better understand the consequences of the coupling between transverse and longitudinal modes, let us first recall the derivation of the standard result \pref{standard}. If one uses the Coulomb gauge and retains in Eq.\ \pref{scouplay} only the coupling to the scalar potential $\phi$, the effect of its integration is a straightforward generalization of Eq.\ \pref{isowg}, provided that one accounts for the anisotropy of the stiffness of Eq.\ \pref{barelay}. The action for the phase would then read:
\be
\lb{isocglay}
S_{\bdnb\cdot\bA=0}^{(ani)}
\simeq 
\frac{\e}{32\pi e^2}\sum_q \left[\frac{\O_m^2 |\bk|^2}{1+\a|\bk|^2}+ \o_{xy}^2 k_{xy}^2+
\o_z^2 k_{z}^2
\right] |\th(q)|^2
\ee
The plasmon dispersion obtained after analytical continuation from the phase propagator in Eq.\ \pref{isocglay} in the limit $\a\ra 0$ is exactly the result \pref{standard}, i.e. 
\bea
\o^2_L(\bk)&=&\frac{1}{|\bk|^2}\left(\o^2_{xy} k_{xy}^2+\o^2_z k_z^2\right)=\nn\\
&=&\o^2_{xy}\frac{k_{xy}^2}{|\bk|^2}+\o^2_z
\frac{k_z^2}{|\bk|^2}=\nn\\
\lb{opmillis}
&=&\o^2_{xy}\sin^2 \eta+\o^2_z\cos^2\eta
\eea
where $\eta$ denotes the angle between the $\bk$ vector and the $z$ axis. Eq.\ \pref{opmillis} is plotted in Fig.\ \ref{fig-RPA} for the values $\o_{xy}=1$ eV and $\o_z=0.05$ eV, which lead to an anisotropy $\g=\o_{xy}/\o_z$ comparable to typical values in cuprates. The  result \pref{opmillis} is analogous to the RPA one derived by including {\em only} the Coulomb potential. This approach, that we will refer to as {\em standard} RPA in what follows, has been adopted in previous work both in the SC 
\cite{randeria_prb00,benfatto_prb01,benfatto_prb04,millis_prr20,vandermarel96,kresin_prb03} and in the normal 
 \cite{fetter_ap74,kresin_prb88,markiewicz_prb08,greco_cp19} state. Within this scheme it is also possible to generalize the expression \pref{opmillis} by retaining the lattice periodicity in the $z$ direction\cite{randeria_prb00,millis_prr20,fetter_ap74,kresin_prb88,markiewicz_prb08,greco_cp19}. As noticed in the introduction, the expression \pref{opmillis} has a singular limit at $\bk=\bf{0}$, since its value depends on the angle $\eta$. As we shall see below, such a singularity is removed by inclusion of the coupling of the phase to the vector potential $\bA$. Finally, for the sake of the following discussion let us also write explicitly the transverse mode obtained in the same approximation, where one neglects the coupling between $\bA_T$ and $\th$. In this case the dispersion of the transverse e.m. wave is simply encoded in the $\bA_T^2$ term of Eq.\ \pref{semtotlay} and it reads:
\be
\lb{otmillis}
\o_T^2(\bk)\equiv
\o_z^2\frac{k^2_{xy}}{|\bk|^2}+\o_{xy}^2\frac{k_z^2}{|\bk|^2} + \ct^2\abs{\bk}^2
\ee
%

%

To fully account for the coupling of the SC phase to both $\bA_L$ and $\bA_T$ let us take advantage of the results of Sec.\ \ref{isotropic} and let us introduce the gauge-invariant phase variables \pref{gipd}. To simplify the notation, we can assume without lack of generality that the in-plane momentum is along the $x$ direction. In this case the $\psi_y$ component decouples from the phase in Eq.\ \pref{scouplay}: it describes a pure massive in-plane transverse mode. The remaining two components are coupled, and with lengthy but straightforward calculations one can derive the generalization of Eq.\ \pref{isopsi}:
\bea
\lb{sani}
 &&S_{ani}[\bpsi]=\frac{\e}{32\pi e^2} \sum_q\nn\\
&&\left[\left(\frac{1+\a k_z^2}{1+\a\abs{\bk}^2}\O_m^2+\o_{xy}^2+\ct^2 k_z^2\right)\abs{\psi_x(q)}^2+\right.\nn\\
&+&\left(\frac{1+\a k_x^2}{1+\a\abs{\bk}^2}\O_m^2+\o_z^2+\ct^2 k_x^2\right)\abs{\psi_z(q)}^2+\nn\\
&-&\left(\frac{\a\O_m^2}
{1+\a\abs{\bk}^2}+\ct^2\right)
k_x k_z\left(\psi_x(q)\psi_z(-q)+c.c. \right)\nn\\
&+&\left.\left(
\O_m^2+\o_{xy}^2+\ct^2|\bk|^2
\right)\abs{\psi_y(q)}^2
\right]
\eea

\begin{figure}[ht]
    \centering
    \includegraphics[width=0.5\textwidth,keepaspectratio]{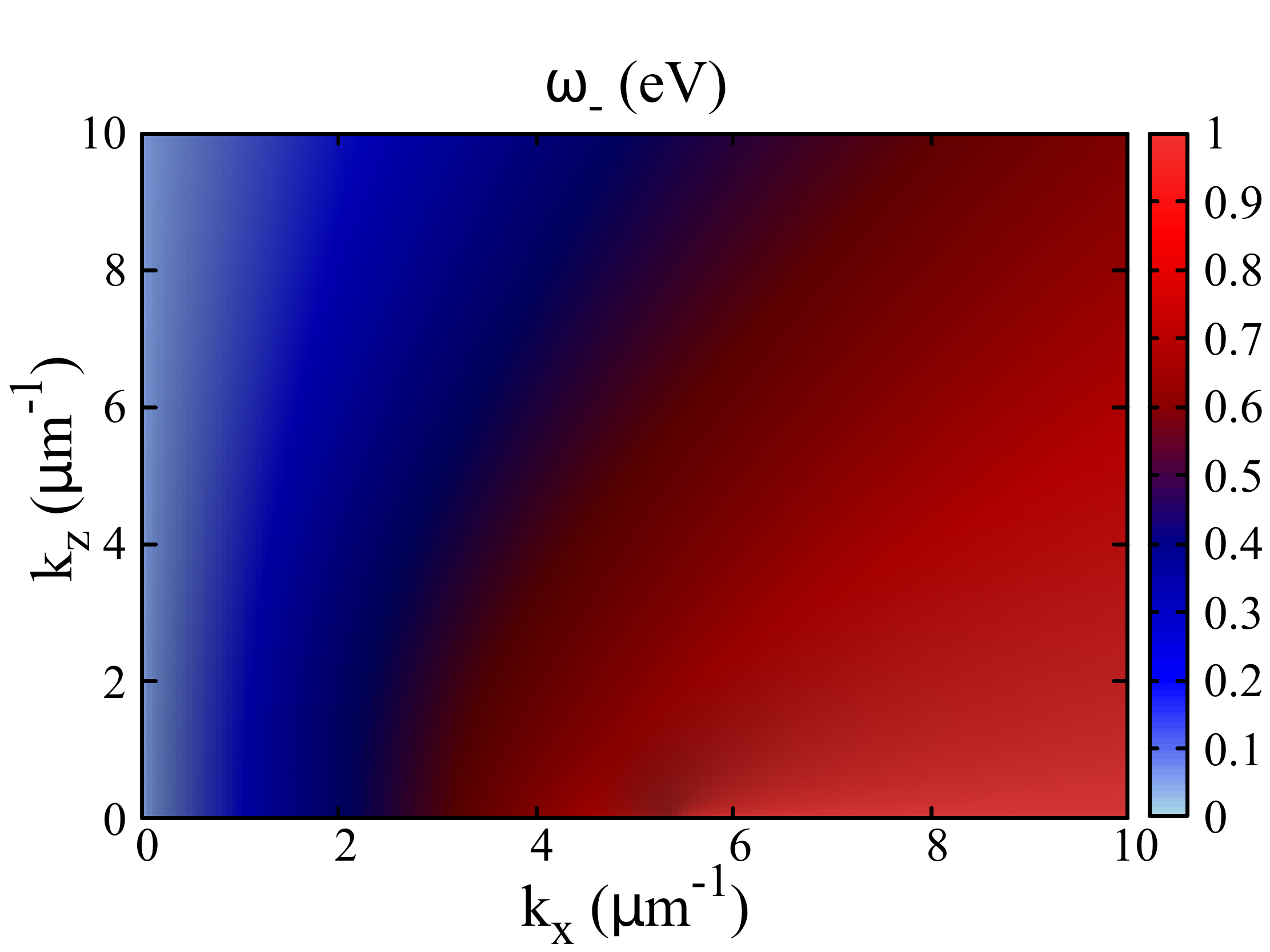}
    \centering
    \includegraphics[width=0.5\textwidth,keepaspectratio]{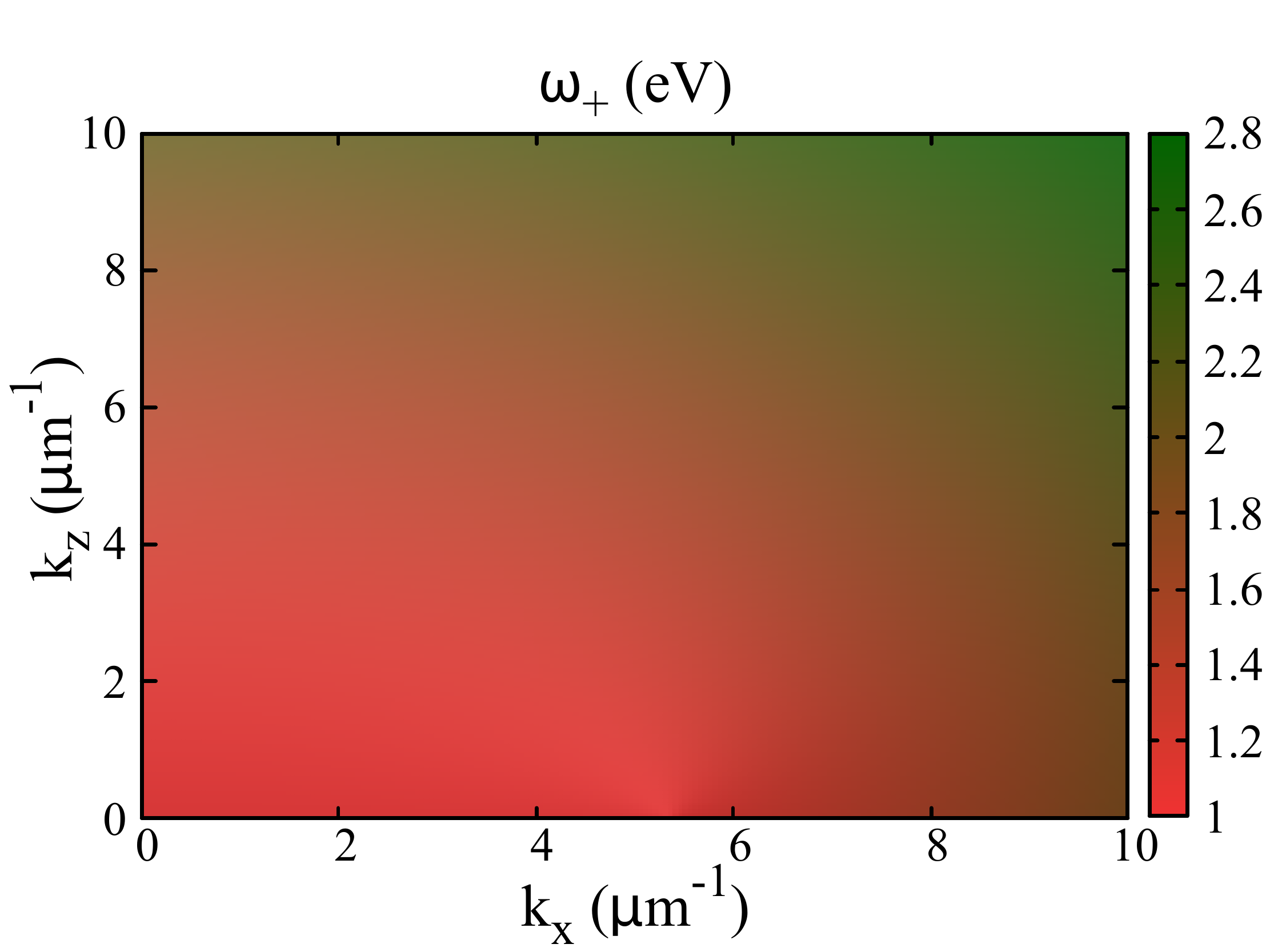}
\caption{ 2D map of the momentum dependence of the two mixed longitudinal-transverse modes $\o_-$ (top) and $\o_+$ (bottom) as given by Eq.\ \pref{ltmodes} in the low-momentum regime $k<k_c$, where $k_c$ is defined by Eq.\ \pref{nrel}. }
\lb{fig-solutions}
\end{figure}
\begin{figure*}[ht]
   \includegraphics[width=1.0\textwidth,keepaspectratio]{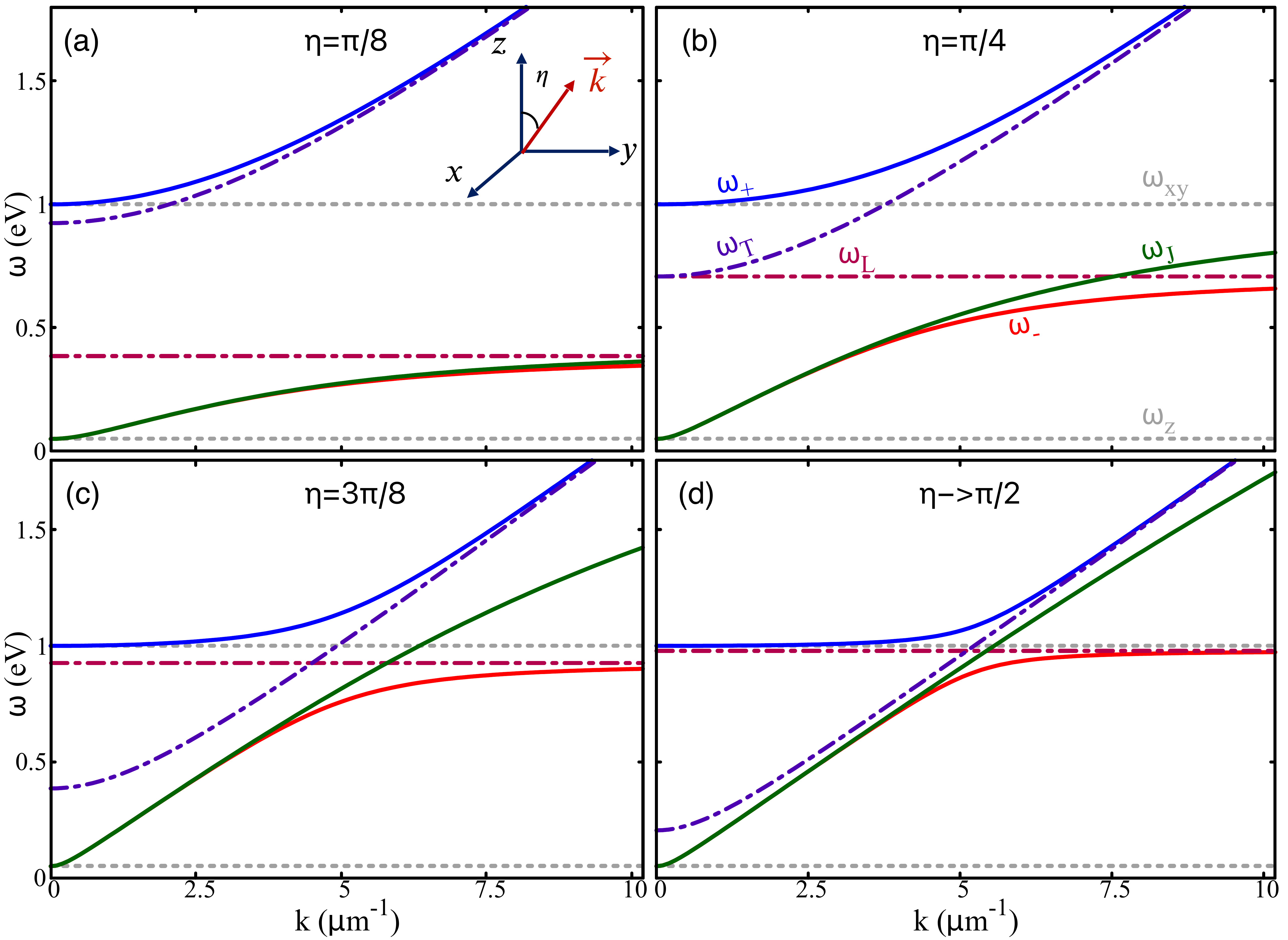}
    \caption{Momentum dependence  of the mixed longitudinal-transverse modes $\o_-$, $\o_+$ (red and blue solid lines, respectively) as a function of $|\bk|$ at selected value of the angle $\eta$ between $\bk$ and the $z$ axis (see inset of panel (a)). We also show for comparison the standard RPA result for the longitudinal $\o_L$ (dot-dashed dark red) and transverse $\o_T$ (dot-dashed dark blue) modes, as given by Eq.s\ \pref{opmillis} and \pref{otmillis}, respectively, and the approximated expression \pref{machida} for $\o_J$ valid around $\o_z$. Dashed grey lines denote the values of the in-plane $\o_{xy}$ and out-of-plane $\o_z$ plasma modes at zero momentum. As before we set $\o_{xy}=1.0$ eV and $\o_z=0.05$ eV and for simplicity we assumed $\epsilon=1$.
    }
    \lb{fig-comparison}
\end{figure*}
We can now compute the propagators $\langle \psi_{\a}(q)\psi_{\b}(-q)\rangle$, whose poles identify the collective-mode excitations. From $\langle |\psi_y(q)|^2\rangle$ one immediately finds the dispersion of the massive in-plane transverse mode $\o^2=\o_{xy}^2+\ct^2|\bk|^2$, that simply reflects the standard isotropic-case result \pref{transv}. The fluctuations of $\psi_x$ and $\psi_z$ are coupled, so the collective modes are given by the determinant of their $2\times 2$ matrix, once the analytic continuation $i\O_m\ra \o+i0^+$ has been performed. The dispersion of the collective excitations is then obtained as solutions of a {\em quartic} characteristic equation, which reads, in the $\a\simeq 0$ limit:

\bea
\lb{chareq}
& &\left(
\o^2-\o_{xy}^2
\right)
\left(
\o^2-\o_z^2
\right)+\nn\\
&&-\ct^2 k_x^2\left(\o^2-\o_{xy}^2\right)
-\ct^2 k_z^2\left(\o^2-\o_z^2\right)
=0
\eea

The solutions of Eq. \pref{chareq} are:

\bea
\lb{ltmodes}
&&\o_{+/-}^2(\bk)=\frac{1}{2}\left(
\o_{xy}^2+\o_z^2+\ct^2|\bk|^2 \pm \right.\nn\\
&\pm &\left. \sqrt{
(\o_{xy}^2-\o_z^2)^2+\ct^4|\bk|^4-2 \ct^2(k_x^2-k_z^2)(\o_{xy}^2-\o_z^2)
}
\right)\nn\\
\eea
\begin{figure*}[ht]
    \centering
    \includegraphics[width=\textwidth]{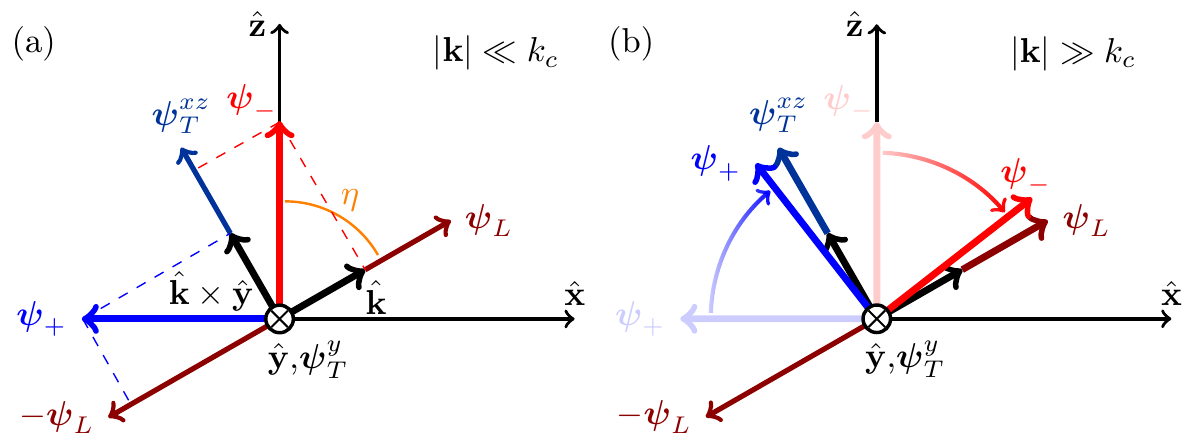}
\caption{{\bf Sketch of the polarization dependence of the mixed modes in a layered superconductors} Polarization dependence of the solutions $\o_-$ and $\o_+$ as given by Eq.\ \pref{freqmodes} at generic momentum.  The black arrows denote  $\hat{\bk}$ and the transverse in-plane direction $\hat{\bk}\times\hat{\by}$. The orange arc denotes the angle $\eta$ between the $z$ axis and the direction $\hat{\bk}$. The longitudinal component $\bpsi_L$ (polarised along $\hat{\bk}$) and the transverse in-plane component $\bpsi_T^{xz}$ (polarized along $\hat{\bk}\times\hat{\by}$) of the g.i. phase difference are, respectively, the dark red and dark blue arrows, while the mixed longitudinal-transverse modes $\bpsi_-$ and $\bpsi_+$ are the red and blue arrows. (a) Low-momentum regime $|\bk|\ll k_c$. Here the $\bpsi_{\mp}$ modes are aligned along the $z$ and $x$ axis, respectively, so at generic $\bk$ they are neither pure longitudinal nor pure transverse. (b) Large-momentum regime $|\bk|\gg k_c$. Here  the mixed mode $\bpsi_-$ reduces to the pure longitudinal mode $\bpsi_L$, while $\bpsi_+$ reduces to the pure transverse mode $\bpsi_T$. The pure transverse mode $\bpsi_T^y$, polarised along the $y$ axis, does not get coupled with the other two degrees of freedom in both cases.}
\lb{fig-polar}
\end{figure*}
Eq.\ \pref{ltmodes} is the first  central result of this paper: it provides the general dispersion of the e.m. excitations of a layered system for any momenta. The two solutions \pref{ltmodes} are shown in Fig.\ \ref{fig-solutions} as two-dimensional maps, while in Fig.\ \ref{fig-comparison} we report the modes as a function of $|\bk|$ for selected value of the angle $\eta$ that $\bk$ forms with the $z$ axis. First of all, we notice that $\o_+$ and $\o_-$ are regular functions as $\bk\ra \bo$. Indeed, assuming that $\o_{xy}>\o_{z}$, as it happens typically in layered materials where planes are weakly coupled, one immediately sees that:
\be
\lb{lowqmodes}
\o_+^2(\bk\ra 0)= \o_{xy}^2, \quad
\o_-^2(\bk\ra 0)=\o_z^2 
\ee
regardless of the direction along which such limit is taken, as it is better shown in Fig. \ref{fig-comparison}. At finite $\bk$ both solutions give an increasing value of the plasma excitation, that is limited above by $\o_{xy}$ for the $\o_-$ solution, while it increases rapidly for the $\o_+$ solution. Such a behavior can be better understood in Fig.\ \ref{fig-comparison} where we plot for comparison also the standard RPA results $\o^2_L$ and $\o^2_T$ derived neglecting the coupling of the phase to the transverse field, see Eq.\ \pref{opmillis} and \pref{otmillis} respectively. Indeed, as $|\bk|$ increases the two solutions \pref{ltmodes} reduce to:
\be
\lb{nonrel}
\begin{cases}
&\o_-(\bk)\simeq
\o_L(\bk),
\\
\\
&\o_+(\bk)\simeq
\o_T(\bk).
\end{cases}
\ee
By closer inspection of Eq.\ \pref{ltmodes} one sees that the crossover among the two regimes occurs around a critical value $k_c$ of the momentum set by the ratio between the plasma-frequency anisotropy and the light velocity:
{\be
\lb{nrel}
k_c=\frac{\sqrt{\o_{xy}^2-\o_z^2}}{\ct}.
\ee
Indeed, as soon as $k\gg k_c$ the square-root term in Eq.\ \pref{ltmodes} can be expanded in powers of the small parameter $k_c/k$ and one easily recovers the two analytical expressions of the purely longitudinal and transverse modes $\o_L$ and $\o_T$}, respectively. To get an idea on the order of magnitude of the momentum scale \pref{nrel}, one should consider that $\hbar c\simeq 0.19 $ eV$\m$m. Thus, considering that in layered materials as e.g. cuprates it is usually $\o_z\ll \o_{xy}$, with $\o_{xy}\simeq 1$ eV, one sees that as soon as $k\gtrsim k_c\sim 5 \, \m $m$^{-1}$ the standard RPA result \pref{opmillis} is recovered, as shown in Fig. \ref{fig-comparison}. As a consequence, for experiments like EELS or RIXS, which measure the plasma dispersion at moment of the order of $1/a\sim 0.1 $\AA$^{-1}$, with $a$ lattice spacing, and energies of the order of the eV, the mixing between longitudinal and transverse modes is not quantitatively relevant. Nonetheless, the discrepancy between $\o_-$ and $\o_L$ becomes crucial in order to understand the radically different description of the low-energy plasma mode provided within the context of non-linear Josephson plasmonic\cite{nori_review10,cavalleri_review}. These experiments are carried out with THz radiation approximately resonant with the low-energy mode $\o\simeq \o_z$ and which, at the boundary with the medium, is polarized along the $z$ axis.  If one considers the solution of Eq.\ \pref{chareq} for $\o\simeq\o_z$, one can approximate $\o^2-\o^2_{xy}\simeq -\o^2_{xy}$ since $\o_{xy}\gg \o_z$. {Physically, this is equivalent to neglect the $(\pd \psi_x/\pd \tau)^2$ term in the first line of Eq.\ \pref{sani} with respect to the $\omega_{xy}^2\psi_x^2$ term, i.e. to assume a stationary in-plane current, that can be a reasonable approximation at the frequency scale of the soft Josephson plasmon.  } From the point of view of the eigenomdes of Eq.\ \pref{sani}, this approximation turns the quartic equation \pref{chareq} into a quadratic one, that can be easily solved leading to Eq.\ \pref{small} mentioned in the introduction, that we can also rewrite as:
\be
\lb{machida}
\o^2_J=\o_z^2\left(1+\frac{\l_z^2 k_x^2}{1+\l_{xy}^2 k_z^2}\right)
\ee
where we introduced the in-plane and out-of-plane penetration depths:
\be
\lb{lambda}
\l_{xy/z}\equiv \frac{\ct}{\o_{xy/z}} 
\ee
As shown in Fig. \ref{fig-comparison}, Eq.\ \pref{machida}  accounts indeed for the correct behaviour of $\o_-$ at energies around $\o_z$ and small momenta. It should be noticed that for optical THz probes the relevant value of the momentum $\bk$ in Eq.\ \pref{machida} is of the order of 10 to 100 cm$^{-1}$, so well below the threshold \pref{nrel}, i.e. in the region where the standard RPA approximation fails. This explains the apparent disagreement between the two expressions usually quoted in the literature for the same problem, studied within the context of different experimental probes. 

%

\subsection{Longitudinal-transverse mixing at generic wavevector}
To gain further insight in the mechanism connecting the general solutions $\o_{\pm}$ to the results obtained with the standard RPA approximation it is instructive to have a closer look to their polarization with respect to $\bk$, in order to understand how the longitudinal-transverse mixing changes around $k_c$. Let us first analyze the low-momentum limit $\bk\ra 0$. In this regime the  eigenvectors corresponding to the two solutions $\omega_{\pm}$ read:
\be
\lb{eigenmodes}
\begin{cases}
& \bpsi_-(\bk)=\frac{ \hat{\bz}+
\frac{\ct^2 k_x k_z}{\o_{xy}^2-\o_z^2}\hat{\bx}}
{\sqrt{1+\frac{\ct^4 k_x^2 k_z^2}{(\o_{xy}^2-\o_z^2)^2}}},
\\
\\
& \bpsi_+(\bk)=\frac{-\hat{\bx}+
\frac{\ct^2 k_x k_z}{\o_{xy}^2-\o_z^2}\hat{\bz}}
{\sqrt{1+\frac{\ct^4 k_x^2 k_z^2}{(\o_{xy}^2-\o_z^2)^2}}}.
\end{cases}
\ee
%
As one clearly sees,  in the $\bk\ra\bo$ limit the $\bpsi_-$ solution is always polarized along $z$, while $\bpsi_+$ is polarized along $x$, see Fig.\ \ref{fig-polar}a, explaining why the plasma modes always reduce to the $\omega_{z/xy}$ values, without the continuum of $\bk\ra 0$ values predicted by the standard RPA, see Fig.\ \ref{fig-RPA}. On the other hand, this also implies that at small momenta the eigenmodes have a mixture of longitudinal and transverse character. This can be better seen by rewriting the two solutions at arbitrary value of the momentum by using explicitly the longitudinal $\bpsi_L$ and transverse $\bpsi_T$ components of the g.i. phase:
\be
\lb{freqmodes}
\begin{cases}
& \bpsi_-(\bk)=\frac{\frac{k_x k_z}{|\bk|^2}\left(\o^2_{xy}-\o_z^2\right)\bpsi_L-\left(\o^2_-(\bk)-\o^2_L(\bk)\right)\bpsi_T^{xz}}{\sqrt{\frac{k_x^2 k_z^2}{|\bk|^4}\left(\o^2_{xy}-\o_z^2\right)^2+\left(\o^2_-(\bk)-\o^2_L(\bk)\right)^2}}
\\
\\
&\bpsi_+(\bk)=\frac{\frac{k_x k_z}{|\bk|^2}\left(\o_{xy}^2-\o_z^2\right)\bpsi_T^{xz}-\left(\o_+(\bk)^2-\o_T(\bk)^2\right)\bpsi_L}{\sqrt{\frac{k_x^2 k_z^2}{|\bk|^4}\left(\o^2_{xy}-\o^2_z\right)^2+\left(\o^2_+(\bk)-\o^2_T(\bk)\right)^2}}
\end{cases}
\ee
%
Since we are setting $\bk$ in the $xz$ plane, one of the transverse components $\bpsi_T^{xz}= (\bpsi_T\times\hat{\by})\times\hat{\by}$ lies in the $xz$ plane along the direction $\hat{\bk}\times\hat{\by}$, while the other $\bpsi_T^y=(\bpsi_T\cdot\hat{\by})\hat{\by}$ is perpendicular to it, see Fig. \ref{fig-polar}. As seen in Eq.\ \pref{sani}, $\psi_y$ is the polarization of the pure transverse mode not involved in the mixing. The orthogonality between $\bpsi_-$ and $\bpsi_+$ is ensured by the fact that $\o_+^2(\bk)+\o_-^2(\bk)=\o_L^2(\bk)+\o_T^2(\bk)$.

Eq.\ \pref{freqmodes} highlights how, at generic value of the momentum, $\o_+$ and $\o_-$ describe two modes with no pure (longitudinal or transverse) character. On the other hand as soon as $k\gg k_c$  and the modes reach the standard RPA value, see Eq.\ \pref{nonrel}, $\bpsi_-$ becomes a purely longitudinal mode while $\bpsi_+$ is a purely transverse one, see  Fig.\ \ref{fig-polar}b, consistently with the expectation that the standard RPA approximation leads to pure longitudinal/transverse modes. 
A remarkable exception to the mixing is provided by the case of propagation completely in plane ($k_z=0$) or completely out-of-plane ($k_x=0$), since in this case the $\psi_x\psi_z$ coupling of Eq.\ \pref{sani} vanishes and one obtains purely longitudinal/transverse modes. For instance when $k_z=0$ the momentum is along $x$, so $\bpsi_+$, which is also aligned along $x$ (see Eq.\ \pref{eigenmodes}), describes a purely longitudinal mode approaching $\omega_{xy}$ as $k_x\ra 0$. Conversely, when $k_x=0$ the momentum is along $z$, so that in this case the longitudinal mode is represented by $\bpsi_-$, and its dispersion approaches $\omega_z$ as $k_z\ra 0$. 
%
%






\subsection{Comparison with the  sine-Gordon equations of motion}\label{sinegordon}
As mentioned in the introduction, the approximated expression \pref{machida} for the small $k$ limit of $\o_-$  has been widely used in the recent  literature\cite{nori_review10,cavalleri_review,nori_natphys06}  to study the dynamics of the Josephson plasmon in the presence of non-linear effects due to intense THz fields. Within this context, it has been derived the equation of motion for the gauge-invariant variable $\psi_z$ of Eq.\ \pref{giphase}. The basic equation reads:
\be
\lb{sineg}
\left(
1-
\l_{xy}^2
 \frac{\pd^2}{\pd z^2}
\right)
\left(
\frac{1}{\o_z^2}
 \frac{\pd^2 \psi_z}{\pd t^2}+\sin \psi_z \right)-\l_z^2
\frac{\pd^2 \psi_z}{\pd x^2}=0,
\ee
where absence of dissipation is assumed, for the sake of simplicity. The explicit appearance of the $\sin \psi_z$ term makes the equation non-linear. {As mentioned above, this result shows how the use of the g.i. phase variable represents  a powerful and elegant way to obtain a straightforward extension of the Gaussian model to the non-linear Josephson model, that represents a crucial step in order to describe non-linear optical effects. On the other hand, the linearized version of Eq.\ \pref{sineg}, obtained  when $\sin \psi_z \simeq \psi_z$, admits a wave-like solution $\psi_z\propto e^{i(\o t - \bk\cdot \br)}$ such that the frequency $\o$ and the momentum $\bk$ satisfy the dispersion relation \pref{machida}. As a consequence, in the linear regime Eq.\ \pref{sineg} provides again an approximation for the $\o_-$ solution at low energy and momenta. It is then interesting to understand how such a non-linear extension to the Josephson model can be obtained starting from the more general formalism developed so far. }

A simple way to see the analogy is to rewrite the general result  \pref{sani} in real space. Consistently with the derivation of Eq.\ \pref{ltmodes}, we shall consider here the $\a=0$ limit. In addition, since Eq.\ \pref{sineg} is written in real time we will convert Matsubara frequencies to real frequencies via analytical continuation and we will replace $\int d\t=\int idt$. The resulting real-time action $\tilde S_{ani}$ then reads:
\bea
\lb{slayer}
\tilde S_{ani}[\psi]&&=
\frac{\e}{32\pi e^2}
\int d t d\bx \Bigg\{ \left(\frac{\pd \psi_x}{\pd t} \right)^2+\left(\frac{\pd \psi_z}{\pd t} \right)^2+\nn\\
-\o_{xy}^2 \psi_x^2-&&\ct^2 \left(\frac{\pd \psi_x }{\pd z}\right)^2+{2}\o_z^2 \cos(\psi_z)-\ct^2 \left(\frac{\pd \psi_z }{\pd x}\right)^2+\nn\\
&&-\ct^2 \left[\psi_x \frac{\pd^2 \psi_z}{\pd x\pd z}+\psi_z \frac{\pd^2 \psi_x }{\pd x\pd z}\right]
\Bigg\}
\eea
Notice that we promoted the $\o_z^2\psi_z^2$ term of Eq.\ \pref{sani} to a cosine-like term $-{2}\o_z^2 \cos \psi_z$, so that interacting terms for the phase are included beyond the Gaussian order. As usual, this is physically motivated by the idea that a Josephson-like coupling set by the out-of-plane phase stiffness $D_z$ exists for the phase in neighboring layers, leading to an effective $XY$ model for the phase degrees of freedom. The main consequence of the presence of a cosine term is that the resulting phase-only model admits naturally a non-linear current along the $z$ direction, $I_z\propto \o_z \sin \psi_z$, that is crucial to account for the non-linear optical response measured at strong THz fields aligned along the $z$ direction in cuprates\cite{cavalleri_natphys16,cavalleri_science18}. As discussed in Ref.\ \cite{benfatto_prb04}, 
the interacting terms for the phase derived microscopically can differ from the one obtained within the simple $XY$ model. Nonetheless, the $XY$ model provides a reasonable starting point to account for non-linear effects in the $z$ direction, as discussed recently in Ref.\ \cite{gabriele_natcomm21}. Finally, to account for the layered structure the gauge-invariant phase variable $\bpsi$ should be promoted to a layer-dependent variable $\bpsi(\br,z=nd)\ra \bpsi_n(\br,t)$, with $\br=(x,y)$, $n$ layer index and $d$ spacing between layers. The final result is completely analogous to Eq.\ \pref{slayer}, provided that one discretizes the integration along $z$ as $\int d\bx\ra {d}\sum_n \int d\br$ and defines a discrete version of the derivative along the $z$ direction, so that $\partial_z f\equiv\frac{1}{d} \partial_nf=\frac{f_{n+1}-f_n}{d}$. We will then retain for simplicity the continuous notation in what follows. 

Once rewritten the action in real space we can obtain the Euler-Lagrange equations of motion by functional derivatives with respect to $\psi_x$ and $\psi_z$. By simple algebra we then obtain two coupled equations which describe the dynamics of the mixed longitudinal-transverse modes:
\bea
\lb{eqmot1}
& &
\frac{\pd^2 \psi_x}{\pd t^2}+\o_{xy}^2\psi_x-\ct^2\frac{\pd^2 \psi_x}{\pd z^2}+
\ct^2\frac{\pd^2 \psi_z}{\pd x\pd z}=0 \\
\lb{eqmot2}
& & \frac{\pd^2 \psi_z}
{\pd t^2}+\o_z^2\sin \psi_z-\ct^2\frac{\pd^2 \psi_z}{\pd x^2}+
\ct^2\frac{\pd^2 \psi_x}{\pd x\pd z}=0
\eea
As one can easily check, when $\sin \psi_z \simeq \psi_z$ one recovers two coupled linear equations which can be solved with wave-like solutions propagating with frequency $\o$ and momenta $\bk$ satisfying the same Eq.\ \pref{chareq} derived above. Thus in the linear regime, as expected, one recovers the same expression \pref{ltmodes} for the $\o_{\pm}$ collective modes derived by the Gaussian action \pref{sani}. On the other hand, if one is interested to study the collective dynamics of $\psi_z$ for frequencies around $\o_z$ one can get an approximate equation of motion by noticing that for $\o\simeq \o_z\ll \o_{xy} $ the first time-derivative term of Eq.\ \pref{eqmot1} is of order $(\o_z/\o_{xy})^2$ as compared to the second one, and can then be neglected. Thus one simply deduces from Eq.\ \pref{eqmot1} that:
\be
\lb{eq1app}
(1-\l_{xy}^2\pd_z^2)\psi_x=-\l_{xy}^2 (\pd_x\pd_z)\psi_z
\ee
where we introduced the penetration depths \pref{lambda}. By applying the $(1-\l_{xy}^2\pd_z^2)$ operator to Eq.\ \pref{eqmot2} and using Eq.\ \pref{eq1app} one obtains exactly Eq.\ \pref{sineg}. In this way, we reconciled all the expression discussed in the previous literature, by clarifying also the limit of validity of the various approximations.

A last comment is in order about the role of the $\a |\bk|^2$ terms, neglected while deriving the $\o_{\pm}$ expressions in Eq.\ \pref{ltmodes}. In the isotropic case, these terms are crucial in order to get the plasmon dispersion, see Eq.\ \pref{omdisp}. Since $\a=\l_D^2$ coincides with the Debye length squared, see Eq.\ \pref{defa}, its effect is negligible as compared to the much larger variations in $\bk$ between $\o_z$ and $\o_{xy}$ already described by Eq.\ \pref{ltmodes}. On the other hand, including these corrections into the general solutions $\o_{\pm}$ is a matter of simple algebra, and in the standard RPA regime one would just recover an additional $\a|\bk|^2$ dispersion into Eq.\ \pref{opmillis}.  

\section{Gauge-invariant density-density response including relativistic corrections}\label{Density}
So far, we investigated the nature of the mixed longitudinal-transverse mode by looking at the dynamics of the gauge-invariant phase variable, and we showed a  failure of the standard RPA approach to capture the $\bk\ra 0$ limit of the plasma dispersion. It is then worth investigating how such generalized plasma modes appear in the physical gauge-invariant density-density response function, which carries information on the longitudinal degrees of freedom of the system.

To clarify the procedure, we first outline the derivation of the density-density response function for an isotropic SC system. The simplest way is to add an auxiliary scalar field $\d\phi$ coupled to the density operator into the microscopic hamiltonian, so that the density-density response can be obtained by functional derivative of the total action with respect to $\d\phi$. In practice, the idea is to derive an action quadratic in $|\d\phi|^2$ after integration out of all the other variables, so that the density-density response function $K$ will be the coefficient of the $|\d\phi|^2$ term in the final action, i.e.
\be
\lb{defk}
K(q)=-\frac{\pd^2 S[\d \phi]}{\pd \d\phi(q) \pd \d \phi(-q)},
\ee
where $S[\d\phi]$ has been obtained by integrating out all other degrees of freedom except then the auxiliary field. In the isotropic case the result is straightforward. Indeed, 
once known the phase-only action \pref{bare} the scalar perturbation $\d\phi$ enters the Gaussian phase-only action via the usual minimal coupling substitution \pref{mincoup}.  One is then left with an effective gaussian model for the variables $\th$, $\d\phi$ and $\phi$, where the action $S_{\d \phi}$ adds to the contribution \pref{sthpsi} computed before. The action $S_{\d\phi}$ contains a bare quadratic term, which accounts for the bare compressibility, and the coupling of the phase with the scalar perturbation. These read: 
\bea
\lb{coupphi}
 S_{\d\phi}[\d\phi,\th]&=&\sum_q
\left[-\frac{\k_0}{2}|\d\phi(q)|^2+\right.\nn\\
&& \left.-\frac{\k_0}{4}\O_m \d\phi(q)\th(-q)
+h.c.
\right]
\eea
As we did before, we can introduce the gauge-invariant phase variable $\bpsi$ and integrate out explicitly $\th$. This leads to a dressing of the bare compressibility term which multiplies $|\d\phi(q)|^2$ in Eq.\ \pref{coupphi} and generates a coupling among $\d\phi$ and $\bpsi$:
\be
\lb{sphipsi}
\begin{split}
S[\bpsi,\d\phi]&=\sum_q\Bigg\{
-\frac{ \k_0 }{2k_{TF}^2} \frac{|\bk|^2
}{1+\alpha|\bk|^2}|\d\phi(q)|^2+\\
+&\frac{\e}{32\pi e^2}\left(\frac{\O_m^2}{1+\a|\bk|^2}+\o_p^2\right)
|\bpsi_L(q)|^2+\\
+&\frac{\k_0}{4k_{TF}^2}\frac{i\O_m|\bk|}{1+\alpha|\bk|^2}
\d\phi(q)|\bpsi_L(-q)|-h.c.\Bigg\}\\
\end{split}.
\ee
As expected, the scalar perturbation couples only to the longitudinal component $\bpsi_L$ of the gauge-invariant phase difference, carrying on the information on the longitudinal modes. Once $\bpsi_L$ is integrated out, we get the fully dressed density-density response function from Eq.\ \pref{defk}. Neglecting the plasmon dispersion (i.e. $\a\simeq 0$) and performing the analytic continuation $i\O_m\ra\o+i\d$ it reads:
\be
\lb{dynlim}
K(q)=\frac{\k_0}{k_{TF}^2}\frac{|\bk|^2\o_p^2}{\o_p^2-(\o+i\delta)^2}.
\ee
 Eq.\ \pref{dynlim} displays a singularity at the plasma frequency $\o_p$. One can easily check that Eq.\ \pref{dynlim} is perfectly equivalent to the standard RPA result $K_{RPA}(q)=\frac{\chi_{0}(q)}{1+V_C(\bk)\chi_{0}(q)}$ obtained in the normal state\cite{nagaosa,fetter}, where $\chi_{0}(q)$ is the Lindhart function for the metal. Indeed, in the limit $\o/k\gg 1$ the Lindhart function can be approximated as $\chi_{0}(q)\simeq -\frac{n}{m}\frac{|\bk|^2}{\o^2}$, that replaced into the standard RPA formula gives exactly Eq. \pref{dynlim}.  Since in this frequency/momentum regime one would expect that the SC susceptibility has the same behavior of the normal-state one, one can conclude that Eq.\ \pref{dynlim} is equivalent to the standard RPA result for the isotropic superconductor. By retaining terms in $\alpha |\bk|^2$ in Eq.\ \pref{sphipsi} one could get also the plasmon dispersion. However, the extension of Eq.\ \pref{dynlim} will not give the correct damping of the plasmon, since it has been derived taking directly the static limit of the density response, i.e. $\kappa_0=\chi_0(\o=0,\bk\ra 0)$. To get the full plasmon spectral function one should retain the full frequency and momentum dependence of the bare density-density response $\chi_0$, in order to recover a plasmon damping when $\omega_p(\bk)$ overlaps the particle-hole continuum\cite{nagaosa,fetter}. 

\begin{figure}[ht]
    \centering
    \includegraphics[width=0.4\textwidth,keepaspectratio]{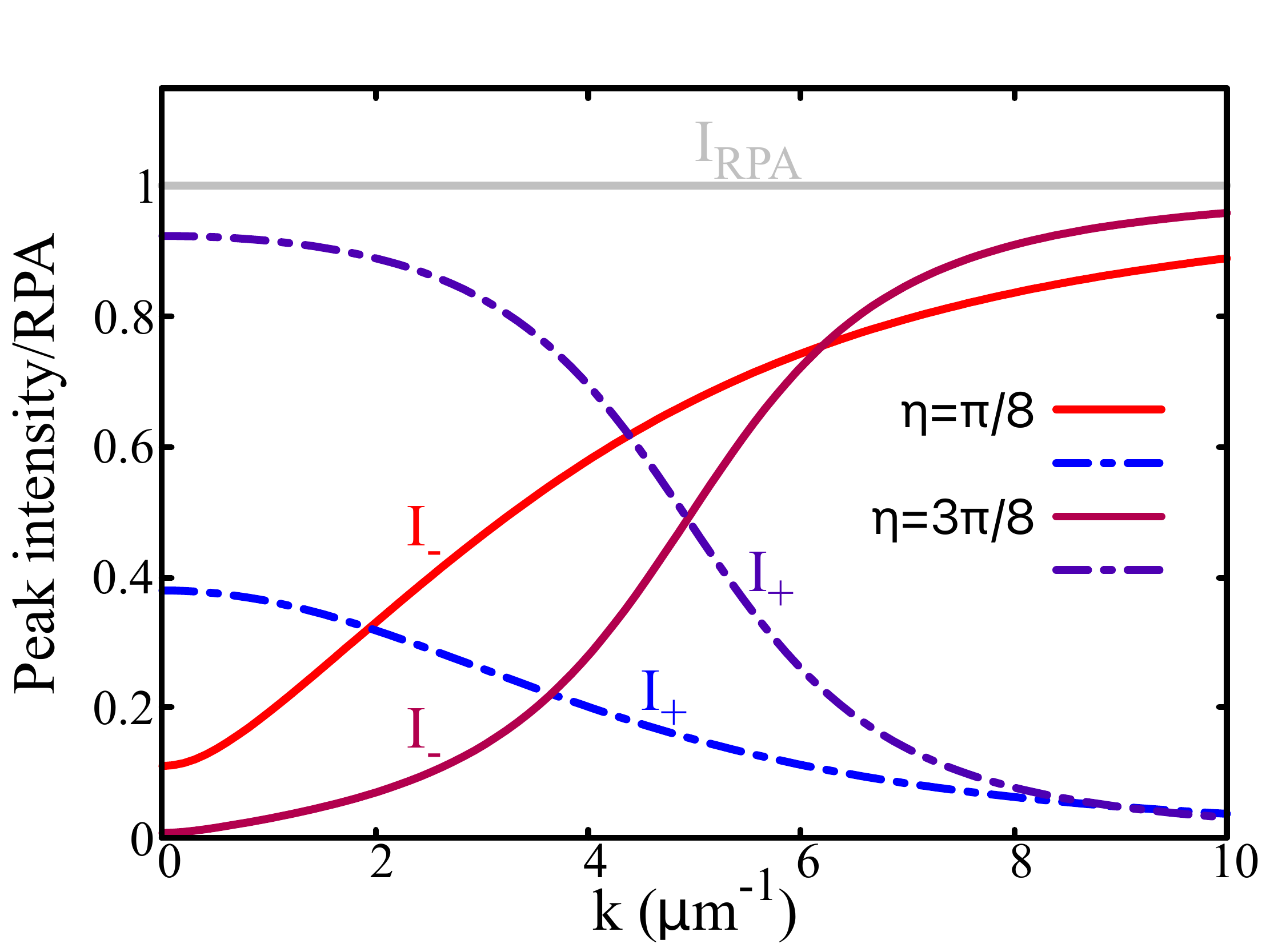}
\caption{Momentum dependence of the spectral weights of the $\o_\pm$ poles in the density-density response. Here we show $I_\pm$ from Eq.\ \pref{ipm} as a function of $|\bk|$ for two fixed values of the angle $\eta$ between $\bk$ and the $z$ axis. In order to compare the results at different angles $\eta$ the intensities$I_\pm$ have been normalized to the intensity of the standard RPA longitudinal mode, i.e. $I_{RPA}=\o_L(\bk)/2$. }
\lb{fig-ipm}
\end{figure}

In order to generalize the previous derivation to the anisotropic case we should introduce, in full analogy with Section III, the in-plane and out-of-plane superfluid stiffness $D_{xy}$ and $D_z$. This leaves unchanged the coupling between $\d\phi$ and $\bpsi_L$ in Eq.\ \pref{sphipsi}, but it modifies the term in $|\bpsi_L|^2$. Once again, the standard RPA result does not account for this effect and it only introduces the anisotropy of the bare Lindhart function. For an anisotropic metal this would be equivalent to approximate $\chi_{0}(q)\simeq -\frac{n}{\o^2}\left( \frac{k_x^2}{m_{xy}}+ \frac{k_z^2}{m_z}\right)$ where the anisotropic in-plane $m_{xy}$ and out-of-plane $m_z$ masses identify the anisotropic plasma frequencies $\o^2_{xy/z}=4\pi e^2 n/m_{xy/z}$. As a consequence, the standard RPA result for the anisotropic case is fully equivalent to Eq.\ \pref{dynlim}, provided that $\o^2_p$ is replaced by its layered version $\o^2_L$ defined in Eq.\ \pref{opmillis}, i.e.
\be
\lb{krpa}
K^{ani}_{RPA}=\frac{\k_0}{k_{TF}^2}\frac{|\bk|^2\o_L(\bk)^2}{\o^2_L(\bk)-(\o+i\delta)^2}.
\ee
As discussed before, we expect that such an approximation fails in the $k\ll k_c$ regime. Indeed, by extending the procedure outlined above for the isotropic case and by integrating out the longitudinal g.i. component $\bpsi_L$ at arbitrary momentum we obtain the following generalization of Eq.\ \pref{dynlim} in the $k\ll k_{TF}$ limit:
%
%
%
\bea
\lb{ddani}
&&K^{(ani)}(q)=\frac{\kappa_0\bk|^2}{k_{TF}^2}
\left(1+\right.\nn\\
&&-\left.\frac{\o^2 \left(\o^2-\o^2_T(\bk)\right)}{\left[(\o+i\delta)^2-
\o_-^2(\bk)\right]\left[(\o+i\delta)^2-
\o_+^2(\bk)\right]}
\right)
\eea
From Eq.\ \pref{ddani}  one sees that the poles of $K^{(ani)}$ are the generalized plasma modes $\o_+$ and $\o_-$ at generic $\bk$. This is consistent with the fact that both have a finite longitudinal projection. Also in this case the full computation of the density spectral function requires a careful evaluation of its imaginary part, that is beyond the scope of the present manuscript. Nonetheless, the result \pref{ddani} already provides us with the direct access to the total spectral weight $I_\pm(\bk)$ allocated in each $\omega_{\pm}(\bk)$ pole as a function of the momentum $\bk$, as given by the imaginary part of Eq.\ \pref{ddani}. These can be readily obtained as:
\bea
\lb{ipm}
I_\pm(\bk)&=&\pm\frac{\o^2_\pm(\bk) (\o^2_\pm(\bk)-\o^2_T(\bk))}{2\o_\pm (\o^2_+-\o^2_-)}.
\eea
The momentum dependence of $I_\pm(\bk)$ is shown in Fig.\ \ref{fig-ipm} for two values of the angle $\eta$ between $\bk$ and the $z$ axis, corresponding to the panels (a) and (c) of Fig.\ \ref{fig-comparison}. Notice that in order to compare the relative spectral weights at a fixed $\bk$ we did not include in Eq.\ \pref{ipm} the overall $|\bk|^2$ factor present in the density-density response \pref{ddani}. As a consequence, in the $k\ll k_c$ regime where the $\o_\pm$ solutions are different from the standard RPA ones $\o_{T/L}$, the upper pole at $\o_+$ has always a larger spectral weight than the lower one $\o_-$, due to the overall $\o^2$ factor in Eq.\ \pref{ddani}. Conversely, in the $k\gg k_c$ limit, where $\o_-$ reduces to the pure RPA longitudinal mode and $\o_+$ to the pure RPA transverse mode (see Eq.\ \pref{nonrel}) only the lower pole at $\o_-$ survives. Indeed, in this regime one sees from Eq.\ \pref{ipm} that the factor $\o^2_+-\o_T^2$ in the numerator of $I_+$ vanishes, while the factor  $\o^2_--\o_+^2\simeq\o^2_--\o_T^2$ at the denominator of $I_-$ cancels out exactly the weighting factor $\o^2_--\o_T^2$ in the numerator, so that $I_-$ approaches the spectral weight of the standard RPA longitudinal mode, which is defined from Eq.\ \pref{krpa} as  $I_{RPA}=\o_L(\bk)/2$. In other words, in this regime the density-density response approaches $K^{ani}_{RPA}$ in Eq.\ \pref{krpa},  obtained in the standard RPA approach. Conversely, at $k\ll k_c$ Eq.\ \pref{ddani} represents a rather unexpected result, i.e. the simultaneous signature in the density response of two generalized plasma modes, located at two well separated energy scales, with the initial predominance of the higher-energy solution $\omega_+$. 


\section{Conclusions}\label{Concl}

In the present manuscript we provided the full description of the generalized plasma modes in bulk layered superconductors, filling the knowledge gap among previous results, that focused on specific regions of the energy/momentum spectrum of the plasmons. The main physical mechanism behind our findings is the anisotropy of the superfluid response in layered superconductors, that leads to an induced superfluid current which is no  more parallel to the electric field. As it is already evident at the level of classical Maxwell's equations (Appendix B), this induces a mixing between transverse and longitudinal components of the  e.m. fields, usually absent in isotropic systems. To rephrase the same effect in an other way, in the anisotropic system density fluctuations are not only generated by a scalar potential but also by a vector potential, and within the context of the superconductor it appears as a finite coupling of the SC phase to the transverse vector potential. Such effects becomes typically relevant for wavevectors that, in strongly anisotropic systems (where i.e. $\o_{xy}\gg \o_z$), are smaller than $k_c\sim \o_{xy}/c$. Since this scales vanishes as $c\ra \infty$ these effects appear as relativistic corrections. At momenta well above the crossover scale the modes acquire an almost pure longitudinal or transverse character, and the anysotropy only manifests in the emergence of acoustic-like plasmon branches reminiscent of the plasma modes in purely 2D systems. 

The above results have been derived by taking advantage of the fact that in a superconductor the plasma modes can be conveniently studied by means of the dynamics of the SC phase of the order parameter. However, previous approaches focused on apparently different models, making it difficult to establish a link between the different expressions for the plasma dispersion discussed in the literature. On this respect, the aim of the present work is twofold. From one side, we provided an  analytical expression for the generalized plasma modes, that smoothly interpolate between the relativistic and the non-relativistic regime. From the other side, we clarified the approximations implicit in the different approaches used in the small-momentum\cite{bulaevskii_prb94,bulaevskii_prb02,machida_prl99,machida_physc00,nori_review10,cavalleri_review,nori_natphys06} or large-momentum\cite{randeria_prb00,benfatto_prb01,benfatto_prb04,millis_prr20,vandermarel96,kresin_prb03} regime, showing that they can be obtained from the same gauge-invariant effective model once specific approximations are done. Finally, to make a direct connection with physically-meausurable quantities, we derived explicitly the density-density response of the layered superconductor, showing how the generalized plasmons are projected out in the physical response in the various energy/momentum regimes. 

Our results, including the methodological aspects, are particularly relevant for future work aimed at investigating e.g. the plasma modes in confined geometries, as it is the case for surface plasmon polaritions\cite{basov-review-polaritons, koppens-review-polaritons}. The recent detection of surface Josephson plasma waves in an ultrathin film of the  high-temperature superconductor La$_{1.85}$Sr$_{0.15}$CuO$_4$ shows that the field is now mature to apply near-field optics, efficiently used so far to investigate van der Waals layered metals\cite{basov-review-polaritons}, to layered superconductors. At the same time, even if standard EELS or RIXS experiments have a rather low momentum resolution, with the lowest accessible momenta of about $\sim 0.1$ of the reciprocal lattice unit, electron energy loss spectroscopy incorporated in a scanning transmission electron microscope (STEM-EELS) and equipped with a monochromator and aberration correctors has a high potential to combine high spatial and energy resolution\cite{pichler_nature19,abajo_natmat21}. As a consequence one can hope in the near future to be able to probe plasma excitations around the crossover scale $k_c\sim 5 \m$m$^{-1}$, where approximate solutions fails, and where both generalized plasma modes give a comparable contribution to the density response, as we showed in Sec.\ \ref{Density}. In addition, at these wavevectors and energies one could explore the possible coupling with other collective modes like phonon, giving rise to hybrid phonon-polaritons modes\cite{koppens-review-polaritons}, involving simultaneously transverse and longitudinal fields. As a final remark, it can be interesting to explore how the transverse-longitudinal mixing affects the plasmon dispersion in the normal state, even though in this case the soft Josephson plasmon is expected to be strongly overdamped by the quasiparticle continuum, especially in correlated systems like cuprates. Such a description can also help understanding how to model the Coulomb interaction among the scattered electrons and the local density of the layered metal in EELS experiments in cuprates, whose interpretation is still debated\cite{mitrano_pnas18,mitrano_prx19,fink_cm21,mitrano_cm21}. Indeed, a robust modelling of the plasma modes in conventional layered superconductors and metals is certainly a prerequisite in order to understand how charge fluctuations manifest instead in a correlated system like cuprates.

\vspace{1cm} {\bf Acknowledgments}
We acknowledge financial support by EU under project MORE-TEM ERC-SYN (grant agreement No 951215).

\appendix
\label{appa}
\section{Phase-only effective action in the path-integral formalism}

We model a generic single-band superconductor via the following grand-canonical hamiltonian:
\be
\lb{ham}
\hat{H}-\m\hat{N}=\sum_{\bk,\s} \xi_{\bk}
\hat{c}^+_{\bk\s}\hat{c}_{\bk\s}
+\hat{H}_I
\ee
where $\xi_{\bk}$ is the band dispersion with respect to the chemical potential $\m$, and the interacting hamiltonian $\hat{H}_I$ reads:
\be
\lb{hamint}
\hat{H}_I=-\frac{U}{N}\sum_\bq
\hat{\Phi}_\D^+(\bq)\hat{\Phi}_\D(\bq)
\ee
$\hat{\Phi}_\D(\bq)\equiv\sum_{\bk\s}\hat{c}_{\bk-\bq/2\s}^+\hat{c}_{\bk+\bq/2,\s}$ being the pairing operator, $U>0$ being the SC coupling constant and $N$ denotes the number of lattice sites. In order to compute thermal averages over the hamiltonian \pref{ham} we use the path integral formulation. Within such framework the the partition function of the system is given by $\mathcal{Z}=\int \mathcal{D}[c,\overline{c}] e^{-S[c,\overline{c}]}$, where $S$ is the imaginary-time action for fermions\cite{nagaosa}:
\be
\begin{split}
&S[c,\overline{c}]=S_0+S_I=\\
&=\int_0^\b d\t
\left[
\sum_{\bk\s}\overline{c}_{\bk\s}
\left(\pd_\t+
\xi_{\bk}
\right)+H_I(\t)
\right]
\end{split}
\ee
$\t=i t$ is the imaginary time, $\b=\frac{1}{T}$ and $c$, $\overline{c}$ are the Grassmann variables associated to the annihilation and creation operators respectively. To obtain the effective action in terms of the order-parameter collective degrees of freedom, the interacting action is decoupled in the particle-particle channel by means of the Hubbard-Stratonovich (HS) transformation by introducing the auxiliary complex field $\D$:
\be
\D(\br)=(\D_0+\d\D(\br))e^{\th (\br)} 
\ee
where $\D_0$ is the mean-field expectation value of the amplitude associated to the SC energy gap, $\d\D$ and $\th$ are amplitude and phase fluctuations. By making an appropriate gauge transformation on the Grassmann fields $c$ and $\overline{c}$ it is possible to make the dependence on the phase $\th$ explicit in the action. One then finds that the HS transform of $S_I$ is phase-independent, while the free contribution $S_0$ {becomes}:
\be
\tilde{S}_0=S_{BCS}+\int d\bx d\t \overline{\Psi}(\bx,\t)
\hat{\S}(\bx,\t)\Psi(\bx,\t)
\ee
where $\Psi$ is the Nambu spinor, defined as the column vector $\left(c_\up, c_\down \right)$, and $S_{BCS}$ is the BCS saddle-point action, where only the mean-field value $\Delta_0$ is the complex field has been included, while fluctuations are contained in $\hat \Sigma$. The BCS Green Function are defined from $S_{BCS}$ as:
\be
\begin{split}
\mathcal{G}_0(\bk,i\o_m)&\equiv -\int_0 ^\b d\t
\langle 
\mathcal{T}\left(\hat{\Psi}_{\bk}(\t)\hat{\Psi}_{\bk}(0)
\right)\rangle e^{i\o_m \t}=\\
&=\frac{i\o_m\hat{\t}_0+\xi_\bk\hat{\t}_3-\D_0\hat{\t}_1}{(i\o_m)^2-E_\bk^2}
\end{split}
\ee
$i\o_m=(2n+1)\pi T$ being Matsubara fermionic frequencies, $E_\bk =\sqrt{\xi_\bk^2+\D_0^2}$ being the quasiparticles energy, $\hat{\t}_i$ being the Pauli matrices. $\hat{\S}$ is the self-energy, which depends, in principle, on both amplitude and phase fluctuations. Nonetheless, as long as one is interested in the low-temperature dynamics of phase fluctuations in layered cuprates, amplitude fluctuations can be neglected\cite{benfatto_prb04}. The self energy then reads:
\be
\lb{seph}
\hat{\S}=\Big\{ \frac{i}{2}\pd_{\t}\th+\frac{1}{8m}\left(\bdnb\th\right)^2\Big\}\hat{\t}_3+\Big\{ \frac{i}{4m} \bdnb\th
\cdot \overleftrightarrow{\nb}
\Big\}\hat{\t}_0
\ee
where $\overleftrightarrow{\nb}\eq \vec{\nb}-\cev{\nb}$. Notice that, according to Goldstone Theorem, the phase $\th$ appears only trough its time and spatial derivatives in Eq.\ \pref{seph}, i.e. no mass term for $\th$ is allowed. Now, since the fermionic variables appear quadratically into {$\tilde{S}_0$}, one can integrate them out. {Such procedure leads to the following effective action for phase fluctuations}:
{\be
\lb{seffexp}
S_{eff}[\th]=\frac{1}{n}\sum_{n=1}^{+\infty}\text{Tr}\Big\{ \left(\hat{\mathcal{G}}_0\hat{\S}\right)^n \Big\},
\ee}
where the trace is computed over both spin and momentum degrees of freedom. In order to study the phase dynamics we compute the effective action at Gaussian level, {i.e. by including terms up to $n=2$}. It reads, in Fourier space:
\bea
\lb{gauss}
S_G[\th]=\frac{1}{8}&\sum_q&
\left[
\O_m^2\L_{\rho\rho}(q)-\bq_a\bq_b\L_{JJ}^{(ab)}+\right.\nn\\
& &\left.+2i\O_m\bq_a\L_{\rho J}^{(a)}
\right]|\th(q)|^2,
\eea
where $q\equiv(i\O_m,\bq)$, {$\O_m=2\pi m T$ being a bosonic Matsubara frequency}, and: 
\be
\L_{\rho\rho}(q)\equiv
\frac{T}{N}\sum_{k}\text{tr}\Big\{
\hat{\mathcal{G}}_0(k+q)\hat{\t}_3
\hat{\mathcal{G}}_0(k)\hat{\t}_3\Big\}
\ee
\be
\L_{\rho J}^a(q)\equiv
\frac{T}{N}\sum_{k}\frac{\bk_a+\frac{\bq_a}{2}}{m}
\text{tr}\Big\{\hat{\mathcal{G}}_0(k+q)\hat{\t}_0
\hat{\mathcal{G}}_0(k)\hat{\t}_3\Big\}
\ee
\bea
&\L_{JJ}^{ab}(q)&\equiv -\frac{n}{m}\d_{ab}+\nn\\
&+&\frac{T}{N}\sum_{k}
\frac{\bk_a+\frac{\bq_a}{2}}{m}
\frac{\bk_b+\frac{\bq_b}{2}}{m}
 \text{tr}\Big\{\hat{\mathcal{G}}_0(k+q)\hat{\t}_0
\hat{\mathcal{G}}_0(k)\hat{\t}_0\Big\}\nn\\
\eea
are the BCS response functions, which contain all the information on the microscopic fermionic degrees of freedom. Again, if one is interested in the low-temperature phase dynamics, the \textit{hydrodynamic limit} of the gaussian action \pref{gauss} is the relevant one. Within such approximation, the BCS bubbles are computed in the static limit $i\O_m=0$, $\bq\ra\bo$, and one finds Eq.\ \pref{bare} of the main text.\\
The formalism developed so far is appropriate for describing neutral superfluids. For the case of superconductors, the e.m. interaction between electrons must be taken into account. This can be achieved via the scalar and vector potentials $\phi$ and $\bA$, which account for internal e.m. degrees of freedom. They can be included into the self-energy \pref{seph} by means of the minimal-coupling substitution, which reads in imaginary time as $\pd_\tau\ra\pd_\tau-e\phi$,  $-i{\bdnb}\ra-i{\bdnb}+\frac{e}{c}\bA$ for the fermionic degrees of freedom. After such a procedure the self energy \pref{seph} undergoes the following modification:
\bea
\lb{sephcharg}
\hat{\S}=\Big\{ &\frac{1}{2}&\left(i\pd_{\t}\th+2e\phi\right)+\frac{1}{8m}\left(\bdnb\th-\frac{2e}{c}\bA\right)^2\Big\}\hat{\t}_3+\nn\\
&+&\Big\{ \frac{i}{4m}\left(\bdnb\th-\frac{2e}{c}\bA\right)
\cdot \overleftrightarrow{\nb}
\Big\}\hat{\t}_0\nn\\
\eea
One immediately sees that Eq.\ \pref{sephcharg} can be  obtained as well by making the substitution \pref{giphase} on the phase degrees of freedom. At this point it is possible to expand the effective action at gaussian level in both the phase and the e.m. potentials: by doing such a procedure and by adding the free-e.m. contribution \pref{sem} one obtains the total action \pref{TotalAction}.\\ External perturbations can be also introduced, in order to compute the response function of the system. For example, in order to compute the density-density response function, one can introduce a scalar perturbation $\d\phi$, which couples with the density operator $\hat{\rho}_{\bq}=\sum_{\bk\s}\hat{c}_{\bk+\bq\s}^+\hat{c}_{\bk\s}$ 
into the microscopic hamiltonian trough the following contribution:
{\be
\lb{deltaphi}
\hat{H}_{\d\phi}=\sum_{\bq,\bk\s} \d\phi(\bq)\hat{c}_{\bk+\bq\s}^+\hat{c}_{\bk\s}.
\ee}
Eq.\ \pref{deltaphi} will give rise to the density insertion $e\d\phi\hat{\t}_3$ into the self energy \pref{sephcharg}. At this point one can compute, trough the usual expansion \pref{seffexp}, the effective action at gaussian level in the phase, the internal e.m. degrees of freedom and the scalar perturbation. Then, once the g.i. variable $\bpsi$ have been introduced and the phase has been integrated out, one is left with Eq.\ \pref{sphipsi}, which describes the coupling of the scalar perturbation $\d\phi$ with the longitudinal degrees of freedom (described by $\bpsi_L$) of the system.

\section{Classical electrodynamics of a layered superconductor}

In this appendix we rephrase the main result of this paper, i.e. the existence of mixed longitudinal-transverse e.m. modes in layered superconductors, within the classical framework of Maxwell's equations. A similar approach has been discussed within the context of the soft Josephson plasmon, see e.g. Ref.\ \cite{bulaevskii_prb94,bulaevskii_prb02,demler_prb20}. Here we show how the set of coupled equations can be solved explicitly at generic frequency and wavevector in order to obtain the two coupled plasma modes provided by Eq.\ \pref{ltmodes}. 
  
We consider a SC system in the absence of external sources, i.e. $\rho_{ext}=0$ and $\bJ_{ext}=\bo$. The superfluid behaviour of the system can be described by means of the first London equation\cite{tinkham}, which relates the internal current $\bJ_{int}$ to the internal electric field $\bE$ in a superconductor:
\be
\lb{londeq}
\frac{\pd \bJ_{int}}{\pd t}=e^2 n_s \hat{m}^{-1} \bE
\ee
where $\hat{m}$ is the effective mass tensor. In isotropic systems it trivially reduces to the scalar effective mass $m$ along an arbitrary direction; on the other hand,  in anisotropic systems it reads 
$$\hat{m}=\begin{pmatrix} m_{xy} & 0 & 0 \\ 0 & m_{xy} & 0 \\ 0 & 0 & m_z \end{pmatrix}$$
where $m_{xy}$ and $m_z$ are the in-plane and the out-of-plane effective masses, respectively.

As it is usually done within the mathematical description of e.m. waves, we take the time derivative of both sides of Biot-Savart law and then substitute Faraday's law, which yields the following equation for the electric field\cite{griffiths}:
\be
\lb{biosav}
\nb^2\bE-\bdnb\left(\bdnb\cdot\bE\right)=
\frac{\e}{c^2}\frac{\pd^2 \bE}{\pd t^2}-
\frac{4\pi}{c^2}\frac{\pd \bJ_{int}}{\pd t}
\ee
We can now get rid of $\bJ_{int}$ by using Eq.\ \pref{londeq}, in order to obtain an equation for the electric field only. Let us introduce the longitudinal $\bE_L=(\hat{\bk}\cdot\bE)\hat{\bk}$ and the transverse $\bE_T=\bE-\bE_L=(\hat{\bk}\times\bE)\times\hat{\bk}$ components of the electric field.  In the isotropic case the longitudinal-transverse decomposition $\bE=\bE_L+\bE_T$ of the total electric field leads to two decoupled equations:
\be
\frac{\pd^2 \bE_L}{\pd t^2}-\o_p^2\bE_L=\bo
\ee
\be
\frac{\pd^2 \bE_T}{\pd t^2}-\ct^2 \nb^2\bE_T-\o_p^2\bE_T=\bo
\ee
{where the renormalized light velocity is defined as $\ct=c/\sqrt{\e}$ as in the main text}. In full analogy with Eq.\ \pref{isopsi}, they describe a longitudinal mode oscillating at $\o=\o_p$ and two degenerate transverse modes propagating at $\o^2=\o_p^2+\ct^2|\bk|^2$, $\o_p$ being the isotropic plasma frequency defined in Eq.\ \pref{omp}. In the anisotropic case such a decomposition for the electric field does not decouple the two equations. The main physical reason is that, due to the tensorial nature of the inverse mass tensor, the induced current ${\bf J}_{int}$ in Eq.\ \pref{londeq} is no more paralell to the electric field.  Let $\hat{x}$ be, as in the main text, the versor parallel to the direction of the in-plane component of the momentum $\bk$. For an anisotropic system Eq.\ \pref{biosav} splits into three equations. One of them describes the in-plane pure transverse component $\bE_T^y=(\hat{\by}\cdot\bE)\hat{\by}$, which reads:
\be
\lb{MaxY}
\left(\o^2-\o_{xy}^2-\ct^2|\bk|^2\right)\bE_T^y=0.
\ee
In full analogy with the behavior of the $\psi_y$ component of the g.i. phase in Eq.\ \pref{sani}, such transverse mode, which is polarized along the $xy$ plane, is not affected by the anisotropy along the out-of-plane direction, so it propagates without coupling with the longitudinal degrees of freedom. On the other hand, the two equations describing the longitudinal mode $\bE_L$ and the transverse component $\bE_T^{xz}=(\hat{\by}\times\bE)\times\hat{\by}=E_T^{xz}(\hat{\bk}\times\hat{\by})$ polarized along the $xz$ plane are coupled. Such equations read, in Fourier space:
\be
\lb{MaxAni}
\begin{cases}
&\left(\o^2-\o_{xy}^2\frac{k_x^2}{|\bk|^2}-\o_z^2\frac{k_z^2}{|\bk|^2}\right)E_L+\\
&+\frac{k_x k_z}{|\bk|^2}\left(\o_{xy}^2-\o_z^2\right)E_T^{xz}=0
\\
\\
&\left(\o^2-\o_z^2\frac{k_x^2}{|\bk|^2}-\o_{xy}^2\frac{k_z^2}{|\bk|^2}-\ct^2|\bk|^2\right)E_T^{xz}+\\
&+\frac{k_x k_z}{|\bk|^2}\left(\o_{xy}^2-\o_z^2\right)E_L=0
\end{cases}
\ee
The non-trivial propagating solutions of the previous equations  are provided by same the solution of the characteristic polynomial \pref{chareq} obtained from the effective action \pref{sani}, leading to the frequencies $\o_{\pm}$ introduced above. The electric fields $\bE_{\pm}$ associated with such modes can be then computed: one finds that, at leading order in $\bk$, they are given exactly by Eq.\ \pref{eigenmodes} of the main text, while at generic momentum they are provided by the general decomposition \pref{freqmodes} in longitudinal and transverse components. Notice that, if the coupling terms $\frac{k_x k_z}{|\bk|^2}(\o_{xy}^2-\o_z^2)E_T^{xz}$ and $\frac{k_x k_z}{|\bk|^2}(\o_{xy}^2-\o_z^2)E_L$ are neglected in both equations (this is a suitable approximation in the non-relativistic regime \pref{nrel}), the pure longitudinal and transverse {standard RPA} modes $\o_L$ and $\o_T$ are recovered.

Lastly, we remark that Eqs.\ \pref{MaxAni} can be recast in a form which is more similar to the one suggested by the action \pref{sani} of the main text, which is $\hat{D}(q)
\bE(q)=\bo$, where the dynamical matrix $\hat{D}$ is defined as follows:
\be
\begin{split}
&\hat{D}(q)\equiv\\
\equiv & \begin{pmatrix}
\o^2-\o_{xy}^2-\ct^2 k_z^2
&
0
&
\ct^2 k_x k_z
\\
0
&
\o^2-\o_{xy}^2-\ct^2|\bk|^2
&
0
\\
\ct^2 k_x k_z
&
0
&
\o^2-\o_z^2-\ct^2 k_x^2
\end{pmatrix}
\end{split}
\lb{DynMatr}
\ee
By using the latter form, the propagating solutions come from the non-triviality condition for the dynamical matrix $\text{Det}\left(\hat{D}(q)\right)=0$. Notice that the matrix $\hat{D}(q)$ and the one associated with Eqs. \pref{MaxY} and \pref{MaxAni} carry the same information, but in two different basis, respectively the standard cartesian basis and the one spanned by the vectors $\hat{\bk}$, $\hat{\by}$ and $\hat{\bk}\times\hat{\by}$. The rotation matrix among the two basis reads $\hat{U}(q)=\frac{1}{|\bk|}\begin{pmatrix} k_x & 0 & k_z \\ 0 & 1 & 0 \\ -k_z & 0 & k_x \end{pmatrix}$. As a consequence the $3\times3$ matrix associated with Eqs. \pref{MaxY} and \pref{MaxAni} can be trivially computed as $\hat{U}\hat{D}\hat{U}^T$.

\bibliography{Literature.bib}

\begin{thebibliography}{54}%
\makeatletter
\providecommand \@ifxundefined [1]{%
 \@ifx{#1\undefined}
}%
\providecommand \@ifnum [1]{%
 \ifnum #1\expandafter \@firstoftwo
 \else \expandafter \@secondoftwo
 \fi
}%
\providecommand \@ifx [1]{%
 \ifx #1\expandafter \@firstoftwo
 \else \expandafter \@secondoftwo
 \fi
}%
\providecommand \natexlab [1]{#1}%
\providecommand \enquote  [1]{``#1''}%
\providecommand \bibnamefont  [1]{#1}%
\providecommand \bibfnamefont [1]{#1}%
\providecommand \citenamefont [1]{#1}%
\providecommand \href@noop [0]{\@secondoftwo}%
\providecommand \href [0]{\begingroup \@sanitize@url \@href}%
\providecommand \@href[1]{\@@startlink{#1}\@@href}%
\providecommand \@@href[1]{\endgroup#1\@@endlink}%
\providecommand \@sanitize@url [0]{\catcode `\\12\catcode `\$12\catcode
  `\&12\catcode `\#12\catcode `\^12\catcode `\_12\catcode `\%12\relax}%
\providecommand \@@startlink[1]{}%
\providecommand \@@endlink[0]{}%
\providecommand \url  [0]{\begingroup\@sanitize@url \@url }%
\providecommand \@url [1]{\endgroup\@href {#1}{\urlprefix }}%
\providecommand \urlprefix  [0]{URL }%
\providecommand \Eprint [0]{\href }%
\providecommand \doibase [0]{http://dx.doi.org/}%
\providecommand \selectlanguage [0]{\@gobble}%
\providecommand \bibinfo  [0]{\@secondoftwo}%
\providecommand \bibfield  [0]{\@secondoftwo}%
\providecommand \translation [1]{[#1]}%
\providecommand \BibitemOpen [0]{}%
\providecommand \bibitemStop [0]{}%
\providecommand \bibitemNoStop [0]{.\EOS\space}%
\providecommand \EOS [0]{\spacefactor3000\relax}%
\providecommand \BibitemShut  [1]{\csname bibitem#1\endcsname}%
\let\auto@bib@innerbib\@empty
\bibitem [{\citenamefont {Maier}(2007)}]{maier}%
  \BibitemOpen
  \bibfield  {author} {\bibinfo {author} {\bibfnamefont {S.}~\bibnamefont
  {Maier}},\ }\href@noop {} {{\selectlanguage {English}\emph {\bibinfo {title}
  {Plasmonics - Fundamentals and Applications}}}}\ (\bibinfo  {publisher}
  {Springer},\ \bibinfo {year} {2007})\BibitemShut {NoStop}%
\bibitem [{\citenamefont {Nozieres}\ and\ \citenamefont {Pines}(1999)}]{pines}%
  \BibitemOpen
  \bibfield  {author} {\bibinfo {author} {\bibfnamefont {P.}~\bibnamefont
  {Nozieres}}\ and\ \bibinfo {author} {\bibfnamefont {D.}~\bibnamefont
  {Pines}},\ }\href {https://books.google.it/books?id=q3wCwaV-gmUC} {\emph
  {\bibinfo {title} {Theory Of Quantum Liquids}}},\ Advanced Books Classics\
  (\bibinfo  {publisher} {Avalon Publishing},\ \bibinfo {year}
  {1999})\BibitemShut {NoStop}%
\bibitem [{\citenamefont {Fetter}\ and\ \citenamefont
  {Walecka}(1971)}]{fetter}%
  \BibitemOpen
  \bibfield  {author} {\bibinfo {author} {\bibfnamefont {A.~L.}\ \bibnamefont
  {Fetter}}\ and\ \bibinfo {author} {\bibfnamefont {J.~D.}\ \bibnamefont
  {Walecka}},\ }\href@noop {} {\emph {\bibinfo {title} {Quantum Theory of
  Many-Particle Systems}}}\ (\bibinfo  {publisher} {McGraw-Hill},\ \bibinfo
  {address} {Boston},\ \bibinfo {year} {1971})\BibitemShut {NoStop}%
\bibitem [{\citenamefont {Low}\ \emph {et~al.}(2017)\citenamefont {Low},
  \citenamefont {Chaves}, \citenamefont {Caldwell}, \citenamefont {Kumar},
  \citenamefont {Fang}, \citenamefont {Avouris}, \citenamefont {Heinz},
  \citenamefont {Guinea}, \citenamefont {Martin-Moreno},\ and\ \citenamefont
  {Koppens}}]{koppens-review-polaritons}%
  \BibitemOpen
  \bibfield  {author} {\bibinfo {author} {\bibfnamefont {T.}~\bibnamefont
  {Low}}, \bibinfo {author} {\bibfnamefont {A.}~\bibnamefont {Chaves}},
  \bibinfo {author} {\bibfnamefont {J.~D.}\ \bibnamefont {Caldwell}}, \bibinfo
  {author} {\bibfnamefont {A.}~\bibnamefont {Kumar}}, \bibinfo {author}
  {\bibfnamefont {N.~X.}\ \bibnamefont {Fang}}, \bibinfo {author}
  {\bibfnamefont {P.}~\bibnamefont {Avouris}}, \bibinfo {author} {\bibfnamefont
  {T.~F.}\ \bibnamefont {Heinz}}, \bibinfo {author} {\bibfnamefont
  {F.}~\bibnamefont {Guinea}}, \bibinfo {author} {\bibfnamefont
  {L.}~\bibnamefont {Martin-Moreno}}, \ and\ \bibinfo {author} {\bibfnamefont
  {F.}~\bibnamefont {Koppens}},\ }\href {\doibase 10.1038/nmat4792} {\bibfield
  {journal} {\bibinfo  {journal} {Nature Materials}\ }\textbf {\bibinfo
  {volume} {16}},\ \bibinfo {pages} {182} (\bibinfo {year} {2017})}\BibitemShut
  {NoStop}%
\bibitem [{\citenamefont {Basov}\ \emph {et~al.}(2016)\citenamefont {Basov},
  \citenamefont {Fogler},\ and\ \citenamefont
  {de~Abajo}}]{basov-review-polaritons}%
  \BibitemOpen
  \bibfield  {author} {\bibinfo {author} {\bibfnamefont {D.~N.}\ \bibnamefont
  {Basov}}, \bibinfo {author} {\bibfnamefont {M.~M.}\ \bibnamefont {Fogler}}, \
  and\ \bibinfo {author} {\bibfnamefont {F.~J.~G.}\ \bibnamefont {de~Abajo}},\
  }\href {\doibase 10.1126/science.aag1992} {\bibfield  {journal} {\bibinfo
  {journal} {Science}\ }\textbf {\bibinfo {volume} {354}},\ \bibinfo {pages}
  {aag1992} (\bibinfo {year} {2016})},\ \Eprint
  {http://arxiv.org/abs/https://www.science.org/doi/pdf/10.1126/science.aag1992}
  {https://www.science.org/doi/pdf/10.1126/science.aag1992} \BibitemShut
  {NoStop}%
\bibitem [{\citenamefont {Nagaosa}\ and\ \citenamefont
  {Heusler}(1999)}]{nagaosa}%
  \BibitemOpen
  \bibfield  {author} {\bibinfo {author} {\bibfnamefont {N.}~\bibnamefont
  {Nagaosa}}\ and\ \bibinfo {author} {\bibfnamefont {S.}~\bibnamefont
  {Heusler}},\ }\href {https://books.google.it/books?id=C9uAXYIlFhMC} {\emph
  {\bibinfo {title} {Quantum Field Theory in Condensed Matter Physics}}},\
  Texts and monographs in physics\ (\bibinfo  {publisher} {Springer},\ \bibinfo
  {year} {1999})\BibitemShut {NoStop}%
\bibitem [{\citenamefont {Anderson}(1958)}]{anderson_pr58}%
  \BibitemOpen
  \bibfield  {author} {\bibinfo {author} {\bibfnamefont {P.~W.}\ \bibnamefont
  {Anderson}},\ }\href {\doibase 10.1103/PhysRev.112.1900} {\bibfield
  {journal} {\bibinfo  {journal} {Phys. Rev.}\ }\textbf {\bibinfo {volume}
  {112}},\ \bibinfo {pages} {1900} (\bibinfo {year} {1958})}\BibitemShut
  {NoStop}%
\bibitem [{\citenamefont {Aitchison}\ \emph {et~al.}(1995)\citenamefont
  {Aitchison}, \citenamefont {Ao}, \citenamefont {Thouless},\ and\
  \citenamefont {Zhu}}]{aitchison_prb95}%
  \BibitemOpen
  \bibfield  {author} {\bibinfo {author} {\bibfnamefont {I.~J.~R.}\
  \bibnamefont {Aitchison}}, \bibinfo {author} {\bibfnamefont {P.}~\bibnamefont
  {Ao}}, \bibinfo {author} {\bibfnamefont {D.~J.}\ \bibnamefont {Thouless}}, \
  and\ \bibinfo {author} {\bibfnamefont {X.-M.}\ \bibnamefont {Zhu}},\ }\href
  {\doibase 10.1103/PhysRevB.51.6531} {\bibfield  {journal} {\bibinfo
  {journal} {Phys. Rev. B}\ }\textbf {\bibinfo {volume} {51}},\ \bibinfo
  {pages} {6531} (\bibinfo {year} {1995})}\BibitemShut {NoStop}%
\bibitem [{\citenamefont {De~Palo}\ \emph {et~al.}(1999)\citenamefont
  {De~Palo}, \citenamefont {Castellani}, \citenamefont {Di~Castro},\ and\
  \citenamefont {Chakraverty}}]{depalo_prb99}%
  \BibitemOpen
  \bibfield  {author} {\bibinfo {author} {\bibfnamefont {S.}~\bibnamefont
  {De~Palo}}, \bibinfo {author} {\bibfnamefont {C.}~\bibnamefont {Castellani}},
  \bibinfo {author} {\bibfnamefont {C.}~\bibnamefont {Di~Castro}}, \ and\
  \bibinfo {author} {\bibfnamefont {B.~K.}\ \bibnamefont {Chakraverty}},\
  }\href {\doibase 10.1103/PhysRevB.60.564} {\bibfield  {journal} {\bibinfo
  {journal} {Phys. Rev. B}\ }\textbf {\bibinfo {volume} {60}},\ \bibinfo
  {pages} {564} (\bibinfo {year} {1999})}\BibitemShut {NoStop}%
\bibitem [{\citenamefont {Paramekanti}\ \emph {et~al.}(2000)\citenamefont
  {Paramekanti}, \citenamefont {Randeria}, \citenamefont {Ramakrishnan},\ and\
  \citenamefont {Mandal}}]{randeria_prb00}%
  \BibitemOpen
  \bibfield  {author} {\bibinfo {author} {\bibfnamefont {A.}~\bibnamefont
  {Paramekanti}}, \bibinfo {author} {\bibfnamefont {M.}~\bibnamefont
  {Randeria}}, \bibinfo {author} {\bibfnamefont {T.~V.}\ \bibnamefont
  {Ramakrishnan}}, \ and\ \bibinfo {author} {\bibfnamefont {S.~S.}\
  \bibnamefont {Mandal}},\ }\href {\doibase 10.1103/PhysRevB.62.6786}
  {\bibfield  {journal} {\bibinfo  {journal} {Phys. Rev. B}\ }\textbf {\bibinfo
  {volume} {62}},\ \bibinfo {pages} {6786} (\bibinfo {year}
  {2000})}\BibitemShut {NoStop}%
\bibitem [{\citenamefont {Benfatto}\ \emph {et~al.}(2001)\citenamefont
  {Benfatto}, \citenamefont {Caprara}, \citenamefont {Castellani},
  \citenamefont {Paramekanti},\ and\ \citenamefont
  {Randeria}}]{benfatto_prb01}%
  \BibitemOpen
  \bibfield  {author} {\bibinfo {author} {\bibfnamefont {L.}~\bibnamefont
  {Benfatto}}, \bibinfo {author} {\bibfnamefont {S.}~\bibnamefont {Caprara}},
  \bibinfo {author} {\bibfnamefont {C.}~\bibnamefont {Castellani}}, \bibinfo
  {author} {\bibfnamefont {A.}~\bibnamefont {Paramekanti}}, \ and\ \bibinfo
  {author} {\bibfnamefont {M.}~\bibnamefont {Randeria}},\ }\href {\doibase
  10.1103/PhysRevB.63.174513} {\bibfield  {journal} {\bibinfo  {journal} {Phys.
  Rev. B}\ }\textbf {\bibinfo {volume} {63}},\ \bibinfo {pages} {174513}
  (\bibinfo {year} {2001})}\BibitemShut {NoStop}%
\bibitem [{\citenamefont {Benfatto}\ \emph {et~al.}(2004)\citenamefont
  {Benfatto}, \citenamefont {Toschi},\ and\ \citenamefont
  {Caprara}}]{benfatto_prb04}%
  \BibitemOpen
  \bibfield  {author} {\bibinfo {author} {\bibfnamefont {L.}~\bibnamefont
  {Benfatto}}, \bibinfo {author} {\bibfnamefont {A.}~\bibnamefont {Toschi}}, \
  and\ \bibinfo {author} {\bibfnamefont {S.}~\bibnamefont {Caprara}},\ }\href
  {\doibase 10.1103/PhysRevB.69.184510} {\bibfield  {journal} {\bibinfo
  {journal} {Phys. Rev. B}\ }\textbf {\bibinfo {volume} {69}},\ \bibinfo
  {pages} {184510} (\bibinfo {year} {2004})}\BibitemShut {NoStop}%
\bibitem [{\citenamefont {Sun}\ \emph {et~al.}(2020)\citenamefont {Sun},
  \citenamefont {Fogler}, \citenamefont {Basov},\ and\ \citenamefont
  {Millis}}]{millis_prr20}%
  \BibitemOpen
  \bibfield  {author} {\bibinfo {author} {\bibfnamefont {Z.}~\bibnamefont
  {Sun}}, \bibinfo {author} {\bibfnamefont {M.~M.}\ \bibnamefont {Fogler}},
  \bibinfo {author} {\bibfnamefont {D.~N.}\ \bibnamefont {Basov}}, \ and\
  \bibinfo {author} {\bibfnamefont {A.~J.}\ \bibnamefont {Millis}},\ }\href
  {\doibase 10.1103/PhysRevResearch.2.023413} {\bibfield  {journal} {\bibinfo
  {journal} {Phys. Rev. Research}\ }\textbf {\bibinfo {volume} {2}},\ \bibinfo
  {pages} {023413} (\bibinfo {year} {2020})}\BibitemShut {NoStop}%
\bibitem [{\citenamefont {Machida}\ \emph {et~al.}(1999)\citenamefont
  {Machida}, \citenamefont {Koyama},\ and\ \citenamefont
  {Tachiki}}]{machida_prl99}%
  \BibitemOpen
  \bibfield  {author} {\bibinfo {author} {\bibfnamefont {M.}~\bibnamefont
  {Machida}}, \bibinfo {author} {\bibfnamefont {T.}~\bibnamefont {Koyama}}, \
  and\ \bibinfo {author} {\bibfnamefont {M.}~\bibnamefont {Tachiki}},\ }\href
  {\doibase 10.1103/PhysRevLett.83.4618} {\bibfield  {journal} {\bibinfo
  {journal} {Phys. Rev. Lett.}\ }\textbf {\bibinfo {volume} {83}},\ \bibinfo
  {pages} {4618} (\bibinfo {year} {1999})}\BibitemShut {NoStop}%
\bibitem [{\citenamefont {Machida}\ \emph {et~al.}(2000)\citenamefont
  {Machida}, \citenamefont {Koyama}, \citenamefont {Tanaka},\ and\
  \citenamefont {Tachiki}}]{machida_physc00}%
  \BibitemOpen
  \bibfield  {author} {\bibinfo {author} {\bibfnamefont {M.}~\bibnamefont
  {Machida}}, \bibinfo {author} {\bibfnamefont {T.}~\bibnamefont {Koyama}},
  \bibinfo {author} {\bibfnamefont {A.}~\bibnamefont {Tanaka}}, \ and\ \bibinfo
  {author} {\bibfnamefont {M.}~\bibnamefont {Tachiki}},\ }\href {\doibase
  https://doi.org/10.1016/S0921-4534(99)00612-7} {\bibfield  {journal}
  {\bibinfo  {journal} {Physica C: Superconductivity}\ }\textbf {\bibinfo
  {volume} {331}},\ \bibinfo {pages} {85 } (\bibinfo {year}
  {2000})}\BibitemShut {NoStop}%
\bibitem [{\citenamefont {Helm}\ and\ \citenamefont
  {Bulaevskii}(2002)}]{bulaevskii_prb02}%
  \BibitemOpen
  \bibfield  {author} {\bibinfo {author} {\bibfnamefont {C.}~\bibnamefont
  {Helm}}\ and\ \bibinfo {author} {\bibfnamefont {L.~N.}\ \bibnamefont
  {Bulaevskii}},\ }\href {\doibase 10.1103/PhysRevB.66.094514} {\bibfield
  {journal} {\bibinfo  {journal} {Phys. Rev. B}\ }\textbf {\bibinfo {volume}
  {66}},\ \bibinfo {pages} {094514} (\bibinfo {year} {2002})}\BibitemShut
  {NoStop}%
\bibitem [{\citenamefont {Savel'ev}\ \emph {et~al.}(2010)\citenamefont
  {Savel'ev}, \citenamefont {Yampol'skii}, \citenamefont {Rakhmanov},\ and\
  \citenamefont {Nori}}]{nori_review10}%
  \BibitemOpen
  \bibfield  {author} {\bibinfo {author} {\bibfnamefont {S.}~\bibnamefont
  {Savel'ev}}, \bibinfo {author} {\bibfnamefont {V.~A.}\ \bibnamefont
  {Yampol'skii}}, \bibinfo {author} {\bibfnamefont {A.~L.}\ \bibnamefont
  {Rakhmanov}}, \ and\ \bibinfo {author} {\bibfnamefont {F.}~\bibnamefont
  {Nori}},\ }\href {\doibase 10.1088/0034-4885/73/2/026501} {\bibfield
  {journal} {\bibinfo  {journal} {Reports on Progress in Physics}\ }\textbf
  {\bibinfo {volume} {73}},\ \bibinfo {pages} {026501} (\bibinfo {year}
  {2010})}\BibitemShut {NoStop}%
\bibitem [{\citenamefont {Laplace}\ and\ \citenamefont
  {Cavalleri}(2016)}]{cavalleri_review}%
  \BibitemOpen
  \bibfield  {author} {\bibinfo {author} {\bibfnamefont {Y.}~\bibnamefont
  {Laplace}}\ and\ \bibinfo {author} {\bibfnamefont {A.}~\bibnamefont
  {Cavalleri}},\ }\href {\doibase 10.1080/23746149.2016.1212671} {\bibfield
  {journal} {\bibinfo  {journal} {Advances in Physics: X}\ }\textbf {\bibinfo
  {volume} {1}},\ \bibinfo {pages} {387} (\bibinfo {year} {2016})}\BibitemShut
  {NoStop}%
\bibitem [{\citenamefont {Savel'ev}\ \emph {et~al.}(2006)\citenamefont
  {Savel'ev}, \citenamefont {Rakhmanov}, \citenamefont {Yampol'skii},\ and\
  \citenamefont {Nori}}]{nori_natphys06}%
  \BibitemOpen
  \bibfield  {author} {\bibinfo {author} {\bibfnamefont {S.}~\bibnamefont
  {Savel'ev}}, \bibinfo {author} {\bibfnamefont {A.~L.}\ \bibnamefont
  {Rakhmanov}}, \bibinfo {author} {\bibfnamefont {V.~A.}\ \bibnamefont
  {Yampol'skii}}, \ and\ \bibinfo {author} {\bibfnamefont {F.}~\bibnamefont
  {Nori}},\ }\href {\doibase 10.1038/nphys358} {\bibfield  {journal} {\bibinfo
  {journal} {Nature Physics}\ }\textbf {\bibinfo {volume} {2}},\ \bibinfo
  {pages} {521} (\bibinfo {year} {2006})}\BibitemShut {NoStop}%
\bibitem [{\citenamefont {van~der Marel}\ and\ \citenamefont
  {Tsvetkov}(1996)}]{vandermarel96}%
  \BibitemOpen
  \bibfield  {author} {\bibinfo {author} {\bibfnamefont {D.}~\bibnamefont
  {van~der Marel}}\ and\ \bibinfo {author} {\bibfnamefont {A.}~\bibnamefont
  {Tsvetkov}},\ }\href {\doibase https://doi.org/10.1007/BF02548125} {\bibfield
   {journal} {\bibinfo  {journal} {Czech. J. of Phys.}\ }\textbf {\bibinfo
  {volume} {46}},\ \bibinfo {pages} {3165} (\bibinfo {year}
  {1996})}\BibitemShut {NoStop}%
\bibitem [{\citenamefont {Bill}\ \emph {et~al.}(2003)\citenamefont {Bill},
  \citenamefont {Morawitz},\ and\ \citenamefont {Kresin}}]{kresin_prb03}%
  \BibitemOpen
  \bibfield  {author} {\bibinfo {author} {\bibfnamefont {A.}~\bibnamefont
  {Bill}}, \bibinfo {author} {\bibfnamefont {H.}~\bibnamefont {Morawitz}}, \
  and\ \bibinfo {author} {\bibfnamefont {V.~Z.}\ \bibnamefont {Kresin}},\
  }\href {\doibase 10.1103/PhysRevB.68.144519} {\bibfield  {journal} {\bibinfo
  {journal} {Phys. Rev. B}\ }\textbf {\bibinfo {volume} {68}},\ \bibinfo
  {pages} {144519} (\bibinfo {year} {2003})}\BibitemShut {NoStop}%
\bibitem [{\citenamefont {Fetter}(1974)}]{fetter_ap74}%
  \BibitemOpen
  \bibfield  {author} {\bibinfo {author} {\bibfnamefont {A.~L.}\ \bibnamefont
  {Fetter}},\ }\href {\doibase https://doi.org/10.1016/0003-4916(74)90397-2}
  {\bibfield  {journal} {\bibinfo  {journal} {Annals of Physics}\ }\textbf
  {\bibinfo {volume} {88}},\ \bibinfo {pages} {1} (\bibinfo {year}
  {1974})}\BibitemShut {NoStop}%
\bibitem [{\citenamefont {Kresin}\ and\ \citenamefont
  {Morawitz}(1988)}]{kresin_prb88}%
  \BibitemOpen
  \bibfield  {author} {\bibinfo {author} {\bibfnamefont {V.~Z.}\ \bibnamefont
  {Kresin}}\ and\ \bibinfo {author} {\bibfnamefont {H.}~\bibnamefont
  {Morawitz}},\ }\href {\doibase 10.1103/PhysRevB.37.7854} {\bibfield
  {journal} {\bibinfo  {journal} {Phys. Rev. B}\ }\textbf {\bibinfo {volume}
  {37}},\ \bibinfo {pages} {7854} (\bibinfo {year} {1988})}\BibitemShut
  {NoStop}%
\bibitem [{\citenamefont {Markiewicz}\ \emph {et~al.}(2008)\citenamefont
  {Markiewicz}, \citenamefont {Hasan},\ and\ \citenamefont
  {Bansil}}]{markiewicz_prb08}%
  \BibitemOpen
  \bibfield  {author} {\bibinfo {author} {\bibfnamefont {R.~S.}\ \bibnamefont
  {Markiewicz}}, \bibinfo {author} {\bibfnamefont {M.~Z.}\ \bibnamefont
  {Hasan}}, \ and\ \bibinfo {author} {\bibfnamefont {A.}~\bibnamefont
  {Bansil}},\ }\href {\doibase 10.1103/PhysRevB.77.094518} {\bibfield
  {journal} {\bibinfo  {journal} {Phys. Rev. B}\ }\textbf {\bibinfo {volume}
  {77}},\ \bibinfo {pages} {094518} (\bibinfo {year} {2008})}\BibitemShut
  {NoStop}%
\bibitem [{\citenamefont {Greco}\ \emph {et~al.}(2019)\citenamefont {Greco},
  \citenamefont {Yamase},\ and\ \citenamefont {Bejas}}]{greco_cp19}%
  \BibitemOpen
  \bibfield  {author} {\bibinfo {author} {\bibfnamefont {A.}~\bibnamefont
  {Greco}}, \bibinfo {author} {\bibfnamefont {H.}~\bibnamefont {Yamase}}, \
  and\ \bibinfo {author} {\bibfnamefont {M.}~\bibnamefont {Bejas}},\ }\href
  {\doibase 10.1038/s42005-018-0099-z} {\bibfield  {journal} {\bibinfo
  {journal} {Communications Physics}\ }\textbf {\bibinfo {volume} {2}},\
  \bibinfo {pages} {3} (\bibinfo {year} {2019})}\BibitemShut {NoStop}%
\bibitem [{\citenamefont {Keimer}\ \emph {et~al.}(2015)\citenamefont {Keimer},
  \citenamefont {Kivelson}, \citenamefont {Norman}, \citenamefont {Uchida},\
  and\ \citenamefont {Zaanen}}]{keimer_review15}%
  \BibitemOpen
  \bibfield  {author} {\bibinfo {author} {\bibfnamefont {B.}~\bibnamefont
  {Keimer}}, \bibinfo {author} {\bibfnamefont {S.~A.}\ \bibnamefont
  {Kivelson}}, \bibinfo {author} {\bibfnamefont {M.~R.}\ \bibnamefont
  {Norman}}, \bibinfo {author} {\bibfnamefont {S.}~\bibnamefont {Uchida}}, \
  and\ \bibinfo {author} {\bibfnamefont {J.}~\bibnamefont {Zaanen}},\ }\href
  {\doibase 10.1038/nature14165} {\bibfield  {journal} {\bibinfo  {journal}
  {Nature}\ }\textbf {\bibinfo {volume} {518}},\ \bibinfo {pages} {179}
  (\bibinfo {year} {2015})}\BibitemShut {NoStop}%
\bibitem [{\citenamefont {Shibauchi}\ \emph {et~al.}(1994)\citenamefont
  {Shibauchi}, \citenamefont {Kitano}, \citenamefont {Uchinokura},
  \citenamefont {Maeda}, \citenamefont {Kimura},\ and\ \citenamefont
  {Kishio}}]{shibauchi_prl94}%
  \BibitemOpen
  \bibfield  {author} {\bibinfo {author} {\bibfnamefont {T.}~\bibnamefont
  {Shibauchi}}, \bibinfo {author} {\bibfnamefont {H.}~\bibnamefont {Kitano}},
  \bibinfo {author} {\bibfnamefont {K.}~\bibnamefont {Uchinokura}}, \bibinfo
  {author} {\bibfnamefont {A.}~\bibnamefont {Maeda}}, \bibinfo {author}
  {\bibfnamefont {T.}~\bibnamefont {Kimura}}, \ and\ \bibinfo {author}
  {\bibfnamefont {K.}~\bibnamefont {Kishio}},\ }\href {\doibase
  10.1103/PhysRevLett.72.2263} {\bibfield  {journal} {\bibinfo  {journal}
  {Phys. Rev. Lett.}\ }\textbf {\bibinfo {volume} {72}},\ \bibinfo {pages}
  {2263} (\bibinfo {year} {1994})}\BibitemShut {NoStop}%
\bibitem [{\citenamefont {Panagopoulos}\ \emph {et~al.}(1996)\citenamefont
  {Panagopoulos}, \citenamefont {Cooper}, \citenamefont {Peacock},
  \citenamefont {Gameson}, \citenamefont {Edwards}, \citenamefont
  {Schmidbauer},\ and\ \citenamefont {Hodby}}]{panagopoulos_prb96}%
  \BibitemOpen
  \bibfield  {author} {\bibinfo {author} {\bibfnamefont {C.}~\bibnamefont
  {Panagopoulos}}, \bibinfo {author} {\bibfnamefont {J.~R.}\ \bibnamefont
  {Cooper}}, \bibinfo {author} {\bibfnamefont {G.~B.}\ \bibnamefont {Peacock}},
  \bibinfo {author} {\bibfnamefont {I.}~\bibnamefont {Gameson}}, \bibinfo
  {author} {\bibfnamefont {P.~P.}\ \bibnamefont {Edwards}}, \bibinfo {author}
  {\bibfnamefont {W.}~\bibnamefont {Schmidbauer}}, \ and\ \bibinfo {author}
  {\bibfnamefont {J.~W.}\ \bibnamefont {Hodby}},\ }\href {\doibase
  10.1103/PhysRevB.53.R2999} {\bibfield  {journal} {\bibinfo  {journal} {Phys.
  Rev. B}\ }\textbf {\bibinfo {volume} {53}},\ \bibinfo {pages} {R2999}
  (\bibinfo {year} {1996})}\BibitemShut {NoStop}%
\bibitem [{\citenamefont {Hosseini}\ \emph {et~al.}(2004)\citenamefont
  {Hosseini}, \citenamefont {Broun}, \citenamefont {Sheehy}, \citenamefont
  {Davis}, \citenamefont {Franz}, \citenamefont {Hardy}, \citenamefont
  {Liang},\ and\ \citenamefont {Bonn}}]{bonn_prl04}%
  \BibitemOpen
  \bibfield  {author} {\bibinfo {author} {\bibfnamefont {A.}~\bibnamefont
  {Hosseini}}, \bibinfo {author} {\bibfnamefont {D.~M.}\ \bibnamefont {Broun}},
  \bibinfo {author} {\bibfnamefont {D.~E.}\ \bibnamefont {Sheehy}}, \bibinfo
  {author} {\bibfnamefont {T.~P.}\ \bibnamefont {Davis}}, \bibinfo {author}
  {\bibfnamefont {M.}~\bibnamefont {Franz}}, \bibinfo {author} {\bibfnamefont
  {W.~N.}\ \bibnamefont {Hardy}}, \bibinfo {author} {\bibfnamefont
  {R.}~\bibnamefont {Liang}}, \ and\ \bibinfo {author} {\bibfnamefont {D.~A.}\
  \bibnamefont {Bonn}},\ }\href {\doibase 10.1103/PhysRevLett.93.107003}
  {\bibfield  {journal} {\bibinfo  {journal} {Phys. Rev. Lett.}\ }\textbf
  {\bibinfo {volume} {93}},\ \bibinfo {pages} {107003} (\bibinfo {year}
  {2004})}\BibitemShut {NoStop}%
\bibitem [{\citenamefont {Tamasaku}\ \emph {et~al.}(1992)\citenamefont
  {Tamasaku}, \citenamefont {Nakamura},\ and\ \citenamefont
  {Uchida}}]{uchida_prl92}%
  \BibitemOpen
  \bibfield  {author} {\bibinfo {author} {\bibfnamefont {K.}~\bibnamefont
  {Tamasaku}}, \bibinfo {author} {\bibfnamefont {Y.}~\bibnamefont {Nakamura}},
  \ and\ \bibinfo {author} {\bibfnamefont {S.}~\bibnamefont {Uchida}},\ }\href
  {\doibase 10.1103/PhysRevLett.69.1455} {\bibfield  {journal} {\bibinfo
  {journal} {Phys. Rev. Lett.}\ }\textbf {\bibinfo {volume} {69}},\ \bibinfo
  {pages} {1455} (\bibinfo {year} {1992})}\BibitemShut {NoStop}%
\bibitem [{\citenamefont {Homes}\ \emph {et~al.}(1993)\citenamefont {Homes},
  \citenamefont {Timusk}, \citenamefont {Liang}, \citenamefont {Bonn},\ and\
  \citenamefont {Hardy}}]{homes_prl93}%
  \BibitemOpen
  \bibfield  {author} {\bibinfo {author} {\bibfnamefont {C.~C.}\ \bibnamefont
  {Homes}}, \bibinfo {author} {\bibfnamefont {T.}~\bibnamefont {Timusk}},
  \bibinfo {author} {\bibfnamefont {R.}~\bibnamefont {Liang}}, \bibinfo
  {author} {\bibfnamefont {D.~A.}\ \bibnamefont {Bonn}}, \ and\ \bibinfo
  {author} {\bibfnamefont {W.~N.}\ \bibnamefont {Hardy}},\ }\href {\doibase
  10.1103/PhysRevLett.71.1645} {\bibfield  {journal} {\bibinfo  {journal}
  {Phys. Rev. Lett.}\ }\textbf {\bibinfo {volume} {71}},\ \bibinfo {pages}
  {1645} (\bibinfo {year} {1993})}\BibitemShut {NoStop}%
\bibitem [{\citenamefont {Kim}\ \emph {et~al.}(1995)\citenamefont {Kim},
  \citenamefont {Somal}, \citenamefont {Czyzyk}, \citenamefont {{van der
  Marel}}, \citenamefont {Wittlin}, \citenamefont {Gerrits}, \citenamefont
  {Duijn}, \citenamefont {Hien},\ and\ \citenamefont
  {Menovsky}}]{kim_physicac95}%
  \BibitemOpen
  \bibfield  {author} {\bibinfo {author} {\bibfnamefont {J.~H.}\ \bibnamefont
  {Kim}}, \bibinfo {author} {\bibfnamefont {H.}~\bibnamefont {Somal}}, \bibinfo
  {author} {\bibfnamefont {M.}~\bibnamefont {Czyzyk}}, \bibinfo {author}
  {\bibfnamefont {D.}~\bibnamefont {{van der Marel}}}, \bibinfo {author}
  {\bibfnamefont {A.}~\bibnamefont {Wittlin}}, \bibinfo {author} {\bibfnamefont
  {A.}~\bibnamefont {Gerrits}}, \bibinfo {author} {\bibfnamefont
  {V.}~\bibnamefont {Duijn}}, \bibinfo {author} {\bibfnamefont
  {N.}~\bibnamefont {Hien}}, \ and\ \bibinfo {author} {\bibfnamefont
  {A.}~\bibnamefont {Menovsky}},\ }\href {\doibase
  https://doi.org/10.1016/0921-4534(95)00198-0} {\bibfield  {journal} {\bibinfo
   {journal} {Physica C: Superconductivity}\ }\textbf {\bibinfo {volume}
  {247}},\ \bibinfo {pages} {297 } (\bibinfo {year} {1995})}\BibitemShut
  {NoStop}%
\bibitem [{\citenamefont {Basov}\ \emph {et~al.}(1994)\citenamefont {Basov},
  \citenamefont {Timusk}, \citenamefont {Dabrowski},\ and\ \citenamefont
  {Jorgensen}}]{basov_prb94}%
  \BibitemOpen
  \bibfield  {author} {\bibinfo {author} {\bibfnamefont {D.~N.}\ \bibnamefont
  {Basov}}, \bibinfo {author} {\bibfnamefont {T.}~\bibnamefont {Timusk}},
  \bibinfo {author} {\bibfnamefont {B.}~\bibnamefont {Dabrowski}}, \ and\
  \bibinfo {author} {\bibfnamefont {J.~D.}\ \bibnamefont {Jorgensen}},\ }\href
  {\doibase 10.1103/PhysRevB.50.3511} {\bibfield  {journal} {\bibinfo
  {journal} {Phys. Rev. B}\ }\textbf {\bibinfo {volume} {50}},\ \bibinfo
  {pages} {3511} (\bibinfo {year} {1994})}\BibitemShut {NoStop}%
\bibitem [{\citenamefont {Bulaevskii}\ \emph {et~al.}(1994)\citenamefont
  {Bulaevskii}, \citenamefont {Zamora}, \citenamefont {Baeriswyl},
  \citenamefont {Beck},\ and\ \citenamefont {Clem}}]{bulaevskii_prb94}%
  \BibitemOpen
  \bibfield  {author} {\bibinfo {author} {\bibfnamefont {L.~N.}\ \bibnamefont
  {Bulaevskii}}, \bibinfo {author} {\bibfnamefont {M.}~\bibnamefont {Zamora}},
  \bibinfo {author} {\bibfnamefont {D.}~\bibnamefont {Baeriswyl}}, \bibinfo
  {author} {\bibfnamefont {H.}~\bibnamefont {Beck}}, \ and\ \bibinfo {author}
  {\bibfnamefont {J.~R.}\ \bibnamefont {Clem}},\ }\href {\doibase
  10.1103/PhysRevB.50.12831} {\bibfield  {journal} {\bibinfo  {journal} {Phys.
  Rev. B}\ }\textbf {\bibinfo {volume} {50}},\ \bibinfo {pages} {12831}
  (\bibinfo {year} {1994})}\BibitemShut {NoStop}%
\bibitem [{\citenamefont {Rajasekaran}\ \emph {et~al.}(2016)\citenamefont
  {Rajasekaran}, \citenamefont {Casandruc}, \citenamefont {Laplace},
  \citenamefont {Nicoletti}, \citenamefont {Gu}, \citenamefont {Clark},
  \citenamefont {Jaksch},\ and\ \citenamefont
  {Cavalleri}}]{cavalleri_natphys16}%
  \BibitemOpen
  \bibfield  {author} {\bibinfo {author} {\bibfnamefont {S.}~\bibnamefont
  {Rajasekaran}}, \bibinfo {author} {\bibfnamefont {E.}~\bibnamefont
  {Casandruc}}, \bibinfo {author} {\bibfnamefont {Y.}~\bibnamefont {Laplace}},
  \bibinfo {author} {\bibfnamefont {D.}~\bibnamefont {Nicoletti}}, \bibinfo
  {author} {\bibfnamefont {G.~D.}\ \bibnamefont {Gu}}, \bibinfo {author}
  {\bibfnamefont {S.~R.}\ \bibnamefont {Clark}}, \bibinfo {author}
  {\bibfnamefont {D.}~\bibnamefont {Jaksch}}, \ and\ \bibinfo {author}
  {\bibfnamefont {A.}~\bibnamefont {Cavalleri}},\ }\href
  {https://doi.org/10.1038/nphys3819} {\bibfield  {journal} {\bibinfo
  {journal} {Nature Physics}\ }\textbf {\bibinfo {volume} {12}},\ \bibinfo
  {pages} {1012} (\bibinfo {year} {2016})}\BibitemShut {NoStop}%
\bibitem [{\citenamefont {Rajasekaran}\ \emph {et~al.}(2018)\citenamefont
  {Rajasekaran}, \citenamefont {Okamoto}, \citenamefont {Mathey}, \citenamefont
  {Fechner}, \citenamefont {Thampy}, \citenamefont {Gu},\ and\ \citenamefont
  {Cavalleri}}]{cavalleri_science18}%
  \BibitemOpen
  \bibfield  {author} {\bibinfo {author} {\bibfnamefont {S.}~\bibnamefont
  {Rajasekaran}}, \bibinfo {author} {\bibfnamefont {J.}~\bibnamefont
  {Okamoto}}, \bibinfo {author} {\bibfnamefont {L.}~\bibnamefont {Mathey}},
  \bibinfo {author} {\bibfnamefont {M.}~\bibnamefont {Fechner}}, \bibinfo
  {author} {\bibfnamefont {V.}~\bibnamefont {Thampy}}, \bibinfo {author}
  {\bibfnamefont {G.~D.}\ \bibnamefont {Gu}}, \ and\ \bibinfo {author}
  {\bibfnamefont {A.}~\bibnamefont {Cavalleri}},\ }\href {\doibase
  10.1126/science.aan3438} {\bibfield  {journal} {\bibinfo  {journal}
  {Science}\ }\textbf {\bibinfo {volume} {359}},\ \bibinfo {pages} {575}
  (\bibinfo {year} {2018})}\BibitemShut {NoStop}%
\bibitem [{\citenamefont {Cremin}\ \emph {et~al.}(2019)\citenamefont {Cremin},
  \citenamefont {Zhang}, \citenamefont {Homes}, \citenamefont {Gu},
  \citenamefont {Sun}, \citenamefont {Fogler}, \citenamefont {Millis},
  \citenamefont {Basov},\ and\ \citenamefont {Averitt}}]{averitt_pnas19}%
  \BibitemOpen
  \bibfield  {author} {\bibinfo {author} {\bibfnamefont {K.~A.}\ \bibnamefont
  {Cremin}}, \bibinfo {author} {\bibfnamefont {J.}~\bibnamefont {Zhang}},
  \bibinfo {author} {\bibfnamefont {C.~C.}\ \bibnamefont {Homes}}, \bibinfo
  {author} {\bibfnamefont {G.~D.}\ \bibnamefont {Gu}}, \bibinfo {author}
  {\bibfnamefont {Z.}~\bibnamefont {Sun}}, \bibinfo {author} {\bibfnamefont
  {M.~M.}\ \bibnamefont {Fogler}}, \bibinfo {author} {\bibfnamefont {A.~J.}\
  \bibnamefont {Millis}}, \bibinfo {author} {\bibfnamefont {D.~N.}\
  \bibnamefont {Basov}}, \ and\ \bibinfo {author} {\bibfnamefont {R.~D.}\
  \bibnamefont {Averitt}},\ }\href {\doibase 10.1073/pnas.1908368116}
  {\bibfield  {journal} {\bibinfo  {journal} {Proceedings of the National
  Academy of Sciences}\ }\textbf {\bibinfo {volume} {116}},\ \bibinfo {pages}
  {19875} (\bibinfo {year} {2019})},\ \Eprint
  {http://arxiv.org/abs/https://www.pnas.org/content/116/40/19875.full.pdf}
  {https://www.pnas.org/content/116/40/19875.full.pdf} \BibitemShut {NoStop}%
\bibitem [{\citenamefont {Michael}\ \emph {et~al.}(2020)\citenamefont
  {Michael}, \citenamefont {von Hoegen}, \citenamefont {Fechner}, \citenamefont
  {F\"orst}, \citenamefont {Cavalleri},\ and\ \citenamefont
  {Demler}}]{demler_prb20}%
  \BibitemOpen
  \bibfield  {author} {\bibinfo {author} {\bibfnamefont {M.~H.}\ \bibnamefont
  {Michael}}, \bibinfo {author} {\bibfnamefont {A.}~\bibnamefont {von Hoegen}},
  \bibinfo {author} {\bibfnamefont {M.}~\bibnamefont {Fechner}}, \bibinfo
  {author} {\bibfnamefont {M.}~\bibnamefont {F\"orst}}, \bibinfo {author}
  {\bibfnamefont {A.}~\bibnamefont {Cavalleri}}, \ and\ \bibinfo {author}
  {\bibfnamefont {E.}~\bibnamefont {Demler}},\ }\href {\doibase
  10.1103/PhysRevB.102.174505} {\bibfield  {journal} {\bibinfo  {journal}
  {Phys. Rev. B}\ }\textbf {\bibinfo {volume} {102}},\ \bibinfo {pages}
  {174505} (\bibinfo {year} {2020})}\BibitemShut {NoStop}%
\bibitem [{\citenamefont {Gabriele}\ \emph {et~al.}(2021)\citenamefont
  {Gabriele}, \citenamefont {Udina},\ and\ \citenamefont
  {Benfatto}}]{gabriele_natcomm21}%
  \BibitemOpen
  \bibfield  {author} {\bibinfo {author} {\bibfnamefont {F.}~\bibnamefont
  {Gabriele}}, \bibinfo {author} {\bibfnamefont {M.}~\bibnamefont {Udina}}, \
  and\ \bibinfo {author} {\bibfnamefont {L.}~\bibnamefont {Benfatto}},\ }\href
  {\doibase 10.1038/s41467-021-21041-6} {\bibfield  {journal} {\bibinfo
  {journal} {Nature Communications}\ }\textbf {\bibinfo {volume} {12}},\
  \bibinfo {pages} {752} (\bibinfo {year} {2021})}\BibitemShut {NoStop}%
\bibitem [{\citenamefont {Stinson}\ \emph {et~al.}(2014)\citenamefont
  {Stinson}, \citenamefont {Wu}, \citenamefont {Jiang}, \citenamefont {Fei},
  \citenamefont {Rodin}, \citenamefont {Chapler}, \citenamefont {McLeod},
  \citenamefont {Castro~Neto}, \citenamefont {Lee}, \citenamefont {Fogler},\
  and\ \citenamefont {Basov}}]{basov_prb14}%
  \BibitemOpen
  \bibfield  {author} {\bibinfo {author} {\bibfnamefont {H.~T.}\ \bibnamefont
  {Stinson}}, \bibinfo {author} {\bibfnamefont {J.~S.}\ \bibnamefont {Wu}},
  \bibinfo {author} {\bibfnamefont {B.~Y.}\ \bibnamefont {Jiang}}, \bibinfo
  {author} {\bibfnamefont {Z.}~\bibnamefont {Fei}}, \bibinfo {author}
  {\bibfnamefont {A.~S.}\ \bibnamefont {Rodin}}, \bibinfo {author}
  {\bibfnamefont {B.~C.}\ \bibnamefont {Chapler}}, \bibinfo {author}
  {\bibfnamefont {A.~S.}\ \bibnamefont {McLeod}}, \bibinfo {author}
  {\bibfnamefont {A.}~\bibnamefont {Castro~Neto}}, \bibinfo {author}
  {\bibfnamefont {Y.~S.}\ \bibnamefont {Lee}}, \bibinfo {author} {\bibfnamefont
  {M.~M.}\ \bibnamefont {Fogler}}, \ and\ \bibinfo {author} {\bibfnamefont
  {D.~N.}\ \bibnamefont {Basov}},\ }\href {\doibase 10.1103/PhysRevB.90.014502}
  {\bibfield  {journal} {\bibinfo  {journal} {Phys. Rev. B}\ }\textbf {\bibinfo
  {volume} {90}},\ \bibinfo {pages} {014502} (\bibinfo {year}
  {2014})}\BibitemShut {NoStop}%
\bibitem [{\citenamefont {Lu}\ \emph {et~al.}(2020)\citenamefont {Lu},
  \citenamefont {Bollinger}, \citenamefont {He}, \citenamefont {Sundling},
  \citenamefont {Bozovic},\ and\ \citenamefont {Gozar}}]{gozar_nqm21}%
  \BibitemOpen
  \bibfield  {author} {\bibinfo {author} {\bibfnamefont {Q.}~\bibnamefont
  {Lu}}, \bibinfo {author} {\bibfnamefont {A.~T.}\ \bibnamefont {Bollinger}},
  \bibinfo {author} {\bibfnamefont {X.}~\bibnamefont {He}}, \bibinfo {author}
  {\bibfnamefont {R.}~\bibnamefont {Sundling}}, \bibinfo {author}
  {\bibfnamefont {I.}~\bibnamefont {Bozovic}}, \ and\ \bibinfo {author}
  {\bibfnamefont {A.}~\bibnamefont {Gozar}},\ }\href {\doibase
  10.1038/s41535-020-00272-8} {\bibfield  {journal} {\bibinfo  {journal} {npj
  Quantum Materials}\ }\textbf {\bibinfo {volume} {5}},\ \bibinfo {pages} {69}
  (\bibinfo {year} {2020})}\BibitemShut {NoStop}%
\bibitem [{\citenamefont {Garc\'{\i}a~de Abajo}(2010)}]{deabajo_review10}%
  \BibitemOpen
  \bibfield  {author} {\bibinfo {author} {\bibfnamefont {F.~J.}\ \bibnamefont
  {Garc\'{\i}a~de Abajo}},\ }\href {\doibase 10.1103/RevModPhys.82.209}
  {\bibfield  {journal} {\bibinfo  {journal} {Rev. Mod. Phys.}\ }\textbf
  {\bibinfo {volume} {82}},\ \bibinfo {pages} {209} (\bibinfo {year}
  {2010})}\BibitemShut {NoStop}%
\bibitem [{\citenamefont {Hepting}\ \emph {et~al.}(2018)\citenamefont
  {Hepting}, \citenamefont {Chaix}, \citenamefont {Huang}, \citenamefont
  {Fumagalli}, \citenamefont {Peng}, \citenamefont {Moritz}, \citenamefont
  {Kummer}, \citenamefont {Brookes}, \citenamefont {Lee}, \citenamefont
  {Hashimoto}, \citenamefont {Sarkar}, \citenamefont {He}, \citenamefont
  {Rotundu}, \citenamefont {Lee}, \citenamefont {Greene}, \citenamefont
  {Braicovich}, \citenamefont {Ghiringhelli}, \citenamefont {Shen},
  \citenamefont {Devereaux},\ and\ \citenamefont {Lee}}]{lee_science18}%
  \BibitemOpen
  \bibfield  {author} {\bibinfo {author} {\bibfnamefont {M.}~\bibnamefont
  {Hepting}}, \bibinfo {author} {\bibfnamefont {L.}~\bibnamefont {Chaix}},
  \bibinfo {author} {\bibfnamefont {E.~W.}\ \bibnamefont {Huang}}, \bibinfo
  {author} {\bibfnamefont {R.}~\bibnamefont {Fumagalli}}, \bibinfo {author}
  {\bibfnamefont {Y.~Y.}\ \bibnamefont {Peng}}, \bibinfo {author}
  {\bibfnamefont {B.}~\bibnamefont {Moritz}}, \bibinfo {author} {\bibfnamefont
  {K.}~\bibnamefont {Kummer}}, \bibinfo {author} {\bibfnamefont {N.~B.}\
  \bibnamefont {Brookes}}, \bibinfo {author} {\bibfnamefont {W.~C.}\
  \bibnamefont {Lee}}, \bibinfo {author} {\bibfnamefont {M.}~\bibnamefont
  {Hashimoto}}, \bibinfo {author} {\bibfnamefont {T.}~\bibnamefont {Sarkar}},
  \bibinfo {author} {\bibfnamefont {J.~F.}\ \bibnamefont {He}}, \bibinfo
  {author} {\bibfnamefont {C.~R.}\ \bibnamefont {Rotundu}}, \bibinfo {author}
  {\bibfnamefont {Y.~S.}\ \bibnamefont {Lee}}, \bibinfo {author} {\bibfnamefont
  {R.~L.}\ \bibnamefont {Greene}}, \bibinfo {author} {\bibfnamefont
  {L.}~\bibnamefont {Braicovich}}, \bibinfo {author} {\bibfnamefont
  {G.}~\bibnamefont {Ghiringhelli}}, \bibinfo {author} {\bibfnamefont {Z.~X.}\
  \bibnamefont {Shen}}, \bibinfo {author} {\bibfnamefont {T.~P.}\ \bibnamefont
  {Devereaux}}, \ and\ \bibinfo {author} {\bibfnamefont {W.~S.}\ \bibnamefont
  {Lee}},\ }\href {\doibase 10.1038/s41586-018-0648-3} {\bibfield  {journal}
  {\bibinfo  {journal} {Nature}\ }\textbf {\bibinfo {volume} {563}},\ \bibinfo
  {pages} {374} (\bibinfo {year} {2018})}\BibitemShut {NoStop}%
\bibitem [{\citenamefont {Lin}\ \emph {et~al.}(2020)\citenamefont {Lin},
  \citenamefont {Yuan}, \citenamefont {Jin}, \citenamefont {Yin}, \citenamefont
  {Li}, \citenamefont {Zhou}, \citenamefont {Lu}, \citenamefont {Dantz},
  \citenamefont {Schmitt}, \citenamefont {Ding}, \citenamefont {Guo},
  \citenamefont {Dean},\ and\ \citenamefont {Liu}}]{liu_nqm20}%
  \BibitemOpen
  \bibfield  {author} {\bibinfo {author} {\bibfnamefont {J.}~\bibnamefont
  {Lin}}, \bibinfo {author} {\bibfnamefont {J.}~\bibnamefont {Yuan}}, \bibinfo
  {author} {\bibfnamefont {K.}~\bibnamefont {Jin}}, \bibinfo {author}
  {\bibfnamefont {Z.}~\bibnamefont {Yin}}, \bibinfo {author} {\bibfnamefont
  {G.}~\bibnamefont {Li}}, \bibinfo {author} {\bibfnamefont {K.-J.}\
  \bibnamefont {Zhou}}, \bibinfo {author} {\bibfnamefont {X.}~\bibnamefont
  {Lu}}, \bibinfo {author} {\bibfnamefont {M.}~\bibnamefont {Dantz}}, \bibinfo
  {author} {\bibfnamefont {T.}~\bibnamefont {Schmitt}}, \bibinfo {author}
  {\bibfnamefont {H.}~\bibnamefont {Ding}}, \bibinfo {author} {\bibfnamefont
  {H.}~\bibnamefont {Guo}}, \bibinfo {author} {\bibfnamefont {M.~P.~M.}\
  \bibnamefont {Dean}}, \ and\ \bibinfo {author} {\bibfnamefont
  {X.}~\bibnamefont {Liu}},\ }\href {\doibase 10.1038/s41535-019-0205-9}
  {\bibfield  {journal} {\bibinfo  {journal} {npj Quantum Materials}\ }\textbf
  {\bibinfo {volume} {5}},\ \bibinfo {pages} {4} (\bibinfo {year}
  {2020})}\BibitemShut {NoStop}%
\bibitem [{\citenamefont {Nag}\ \emph {et~al.}(2020)\citenamefont {Nag},
  \citenamefont {Zhu}, \citenamefont {Bejas}, \citenamefont {Li}, \citenamefont
  {Robarts}, \citenamefont {Yamase}, \citenamefont {Petsch}, \citenamefont
  {Song}, \citenamefont {Eisaki}, \citenamefont {Walters}, \citenamefont
  {Garc\'{\i}a-Fern\'andez}, \citenamefont {Greco}, \citenamefont {Hayden},\
  and\ \citenamefont {Zhou}}]{zhou_prl20}%
  \BibitemOpen
  \bibfield  {author} {\bibinfo {author} {\bibfnamefont {A.}~\bibnamefont
  {Nag}}, \bibinfo {author} {\bibfnamefont {M.}~\bibnamefont {Zhu}}, \bibinfo
  {author} {\bibfnamefont {M.}~\bibnamefont {Bejas}}, \bibinfo {author}
  {\bibfnamefont {J.}~\bibnamefont {Li}}, \bibinfo {author} {\bibfnamefont
  {H.~C.}\ \bibnamefont {Robarts}}, \bibinfo {author} {\bibfnamefont
  {H.}~\bibnamefont {Yamase}}, \bibinfo {author} {\bibfnamefont {A.~N.}\
  \bibnamefont {Petsch}}, \bibinfo {author} {\bibfnamefont {D.}~\bibnamefont
  {Song}}, \bibinfo {author} {\bibfnamefont {H.}~\bibnamefont {Eisaki}},
  \bibinfo {author} {\bibfnamefont {A.~C.}\ \bibnamefont {Walters}}, \bibinfo
  {author} {\bibfnamefont {M.}~\bibnamefont {Garc\'{\i}a-Fern\'andez}},
  \bibinfo {author} {\bibfnamefont {A.}~\bibnamefont {Greco}}, \bibinfo
  {author} {\bibfnamefont {S.~M.}\ \bibnamefont {Hayden}}, \ and\ \bibinfo
  {author} {\bibfnamefont {K.-J.}\ \bibnamefont {Zhou}},\ }\href {\doibase
  10.1103/PhysRevLett.125.257002} {\bibfield  {journal} {\bibinfo  {journal}
  {Phys. Rev. Lett.}\ }\textbf {\bibinfo {volume} {125}},\ \bibinfo {pages}
  {257002} (\bibinfo {year} {2020})}\BibitemShut {NoStop}%
\bibitem [{\citenamefont {Mitrano}\ \emph {et~al.}(2018)\citenamefont
  {Mitrano}, \citenamefont {Husain}, \citenamefont {Vig}, \citenamefont
  {Kogar}, \citenamefont {Rak}, \citenamefont {Rubeck}, \citenamefont
  {Schmalian}, \citenamefont {Uchoa}, \citenamefont {Schneeloch}, \citenamefont
  {Zhong}, \citenamefont {Gu},\ and\ \citenamefont
  {Abbamonte}}]{mitrano_pnas18}%
  \BibitemOpen
  \bibfield  {author} {\bibinfo {author} {\bibfnamefont {M.}~\bibnamefont
  {Mitrano}}, \bibinfo {author} {\bibfnamefont {A.~A.}\ \bibnamefont {Husain}},
  \bibinfo {author} {\bibfnamefont {S.}~\bibnamefont {Vig}}, \bibinfo {author}
  {\bibfnamefont {A.}~\bibnamefont {Kogar}}, \bibinfo {author} {\bibfnamefont
  {M.~S.}\ \bibnamefont {Rak}}, \bibinfo {author} {\bibfnamefont {S.~I.}\
  \bibnamefont {Rubeck}}, \bibinfo {author} {\bibfnamefont {J.}~\bibnamefont
  {Schmalian}}, \bibinfo {author} {\bibfnamefont {B.}~\bibnamefont {Uchoa}},
  \bibinfo {author} {\bibfnamefont {J.}~\bibnamefont {Schneeloch}}, \bibinfo
  {author} {\bibfnamefont {R.}~\bibnamefont {Zhong}}, \bibinfo {author}
  {\bibfnamefont {G.~D.}\ \bibnamefont {Gu}}, \ and\ \bibinfo {author}
  {\bibfnamefont {P.}~\bibnamefont {Abbamonte}},\ }\href {\doibase
  10.1073/pnas.1721495115} {\bibfield  {journal} {\bibinfo  {journal}
  {Proceedings of the National Academy of Sciences}\ }\textbf {\bibinfo
  {volume} {115}},\ \bibinfo {pages} {5392} (\bibinfo {year} {2018})},\ \Eprint
  {http://arxiv.org/abs/https://www.pnas.org/content/115/21/5392.full.pdf}
  {https://www.pnas.org/content/115/21/5392.full.pdf} \BibitemShut {NoStop}%
\bibitem [{\citenamefont {Husain}\ \emph {et~al.}(2019)\citenamefont {Husain},
  \citenamefont {Mitrano}, \citenamefont {Rak}, \citenamefont {Rubeck},
  \citenamefont {Uchoa}, \citenamefont {March}, \citenamefont {Dwyer},
  \citenamefont {Schneeloch}, \citenamefont {Zhong}, \citenamefont {Gu},\ and\
  \citenamefont {Abbamonte}}]{mitrano_prx19}%
  \BibitemOpen
  \bibfield  {author} {\bibinfo {author} {\bibfnamefont {A.~A.}\ \bibnamefont
  {Husain}}, \bibinfo {author} {\bibfnamefont {M.}~\bibnamefont {Mitrano}},
  \bibinfo {author} {\bibfnamefont {M.~S.}\ \bibnamefont {Rak}}, \bibinfo
  {author} {\bibfnamefont {S.}~\bibnamefont {Rubeck}}, \bibinfo {author}
  {\bibfnamefont {B.}~\bibnamefont {Uchoa}}, \bibinfo {author} {\bibfnamefont
  {K.}~\bibnamefont {March}}, \bibinfo {author} {\bibfnamefont
  {C.}~\bibnamefont {Dwyer}}, \bibinfo {author} {\bibfnamefont
  {J.}~\bibnamefont {Schneeloch}}, \bibinfo {author} {\bibfnamefont
  {R.}~\bibnamefont {Zhong}}, \bibinfo {author} {\bibfnamefont {G.~D.}\
  \bibnamefont {Gu}}, \ and\ \bibinfo {author} {\bibfnamefont {P.}~\bibnamefont
  {Abbamonte}},\ }\href {\doibase 10.1103/PhysRevX.9.041062} {\bibfield
  {journal} {\bibinfo  {journal} {Phys. Rev. X}\ }\textbf {\bibinfo {volume}
  {9}},\ \bibinfo {pages} {041062} (\bibinfo {year} {2019})}\BibitemShut
  {NoStop}%
\bibitem [{\citenamefont {Fink}(2021)}]{fink_cm21}%
  \BibitemOpen
  \bibfield  {author} {\bibinfo {author} {\bibfnamefont {J.}~\bibnamefont
  {Fink}},\ }\href@noop {} {}\bibinfo {howpublished} {arXiv:2103.10268}
  (\bibinfo {year} {2021})\BibitemShut {NoStop}%
\bibitem [{\citenamefont {et~al.}(2019)}]{mitrano_cm21}%
  \BibitemOpen
  \bibfield  {author} {\bibinfo {author} {\bibfnamefont {A.~H.}\ \bibnamefont
  {et~al.}},\ }\href@noop {} {}\bibinfo {howpublished} {arXiv:2106.03301}
  (\bibinfo {year} {2019})\BibitemShut {NoStop}%
\bibitem [{\citenamefont {Anderson}(1963)}]{anderson_pr63}%
  \BibitemOpen
  \bibfield  {author} {\bibinfo {author} {\bibfnamefont {P.~W.}\ \bibnamefont
  {Anderson}},\ }\href {\doibase 10.1103/PhysRev.130.439} {\bibfield  {journal}
  {\bibinfo  {journal} {Phys. Rev.}\ }\textbf {\bibinfo {volume} {130}},\
  \bibinfo {pages} {439} (\bibinfo {year} {1963})}\BibitemShut {NoStop}%
\bibitem [{\citenamefont {Senga}\ \emph {et~al.}(2019)\citenamefont {Senga},
  \citenamefont {Suenaga}, \citenamefont {Barone}, \citenamefont {Morishita},
  \citenamefont {Mauri},\ and\ \citenamefont {Pichler}}]{pichler_nature19}%
  \BibitemOpen
  \bibfield  {author} {\bibinfo {author} {\bibfnamefont {R.}~\bibnamefont
  {Senga}}, \bibinfo {author} {\bibfnamefont {K.}~\bibnamefont {Suenaga}},
  \bibinfo {author} {\bibfnamefont {P.}~\bibnamefont {Barone}}, \bibinfo
  {author} {\bibfnamefont {S.}~\bibnamefont {Morishita}}, \bibinfo {author}
  {\bibfnamefont {F.}~\bibnamefont {Mauri}}, \ and\ \bibinfo {author}
  {\bibfnamefont {T.}~\bibnamefont {Pichler}},\ }\href {\doibase
  10.1038/s41586-019-1477-8} {\bibfield  {journal} {\bibinfo  {journal}
  {Nature}\ }\textbf {\bibinfo {volume} {573}},\ \bibinfo {pages} {247}
  (\bibinfo {year} {2019})}\BibitemShut {NoStop}%
\bibitem [{\citenamefont {Li}\ \emph {et~al.}(2021)\citenamefont {Li},
  \citenamefont {Guo}, \citenamefont {Yang}, \citenamefont {Qi}, \citenamefont
  {Qiao}, \citenamefont {Li}, \citenamefont {Shi}, \citenamefont {Li},
  \citenamefont {Liu}, \citenamefont {Xu}, \citenamefont {Liu}, \citenamefont
  {Garc{\'\i}a~de Abajo}, \citenamefont {Dai}, \citenamefont {Wang},\ and\
  \citenamefont {Gao}}]{abajo_natmat21}%
  \BibitemOpen
  \bibfield  {author} {\bibinfo {author} {\bibfnamefont {N.}~\bibnamefont
  {Li}}, \bibinfo {author} {\bibfnamefont {X.}~\bibnamefont {Guo}}, \bibinfo
  {author} {\bibfnamefont {X.}~\bibnamefont {Yang}}, \bibinfo {author}
  {\bibfnamefont {R.}~\bibnamefont {Qi}}, \bibinfo {author} {\bibfnamefont
  {T.}~\bibnamefont {Qiao}}, \bibinfo {author} {\bibfnamefont {Y.}~\bibnamefont
  {Li}}, \bibinfo {author} {\bibfnamefont {R.}~\bibnamefont {Shi}}, \bibinfo
  {author} {\bibfnamefont {Y.}~\bibnamefont {Li}}, \bibinfo {author}
  {\bibfnamefont {K.}~\bibnamefont {Liu}}, \bibinfo {author} {\bibfnamefont
  {Z.}~\bibnamefont {Xu}}, \bibinfo {author} {\bibfnamefont {L.}~\bibnamefont
  {Liu}}, \bibinfo {author} {\bibfnamefont {F.~J.}\ \bibnamefont
  {Garc{\'\i}a~de Abajo}}, \bibinfo {author} {\bibfnamefont {Q.}~\bibnamefont
  {Dai}}, \bibinfo {author} {\bibfnamefont {E.-G.}\ \bibnamefont {Wang}}, \
  and\ \bibinfo {author} {\bibfnamefont {P.}~\bibnamefont {Gao}},\ }\href
  {\doibase 10.1038/s41563-020-0763-z} {\bibfield  {journal} {\bibinfo
  {journal} {Nature Materials}\ }\textbf {\bibinfo {volume} {20}},\ \bibinfo
  {pages} {43} (\bibinfo {year} {2021})}\BibitemShut {NoStop}%
\bibitem [{\citenamefont {Tinkham}(2004)}]{tinkham}%
  \BibitemOpen
  \bibfield  {author} {\bibinfo {author} {\bibfnamefont {M.}~\bibnamefont
  {Tinkham}},\ }\href@noop {} {\emph {\bibinfo {title} {Introduction to
  superconductivity}}}\ (\bibinfo  {publisher} {Courier Corporation},\ \bibinfo
  {year} {2004})\BibitemShut {NoStop}%
\bibitem [{\citenamefont {Griffiths}(2005)}]{griffiths}%
  \BibitemOpen
  \bibfield  {author} {\bibinfo {author} {\bibfnamefont {D.~J.}\ \bibnamefont
  {Griffiths}},\ }\href@noop {} {\enquote {\bibinfo {title} {Introduction to
  electrodynamics},}\ } (\bibinfo {year} {2005})\BibitemShut {NoStop}%
\end{thebibliography}%

\end{document}